# The WFIRST Galaxy Survey Exposure Time Calculator

**Christopher M. Hirata[1]*, Neil Gehrels[2], Jean-Paul Kneib[3], Jeffrey Kruk[4], Jason Rhodes[5], Yun Wang[6], and Julien Zoubian[3]**

[1]*California Institute of Technology M/C 350-17, Pasadena, California 91125, USA*
[2]*NASA Goddard Space Flight Center, Code 661, Greenbelt, Maryland 20771, USA*
[3]*Laboratoire d'Astrophysique de Marseille, CNRS-Université Aix-Marseille, 38 rue F. Joliot-Curie 13388 Marseille Cedex 13, France*
[4]*NASA Goddard Space Flight Center, Code 665, Greenbelt, Maryland 20771, USA*
[5]*Jet Propulsion Laboratory M/S 169-506, 4800 Oak Grove Drive, Pasadena, California, 91109, USA*
[6]*Department of Physics and Astronomy, University of Oklahoma, 440 W Brooks Street, Norman, Oklahoma 73019, USA*

\* e-mail: chirata@tapir.caltech.edu

**April 20, 2012**

This document describes the exposure time calculator for the Wide-Field Infrared Survey Telescope (WFIRST) high-latitude survey. The calculator works in both imaging and spectroscopic modes. In addition to the standard ETC functions (e.g. background and S/N determination), the calculator integrates over the galaxy population and forecasts the density and redshift distribution of galaxy shapes usable for weak lensing (in imaging mode) and the detected emission lines (in spectroscopic mode). The source code is made available for public use.

# Space-Based Galaxy Survey Exposure Time Calculator
# Version 10
# User Manual & Code Description

## Table of Contents

# 1. Overview

This ETC is intended for use in optimizing the WFIRST hardware and observing strategy. It runs in three different modes: WL (shape measurement), BAO (galaxy redshifts), and PZCAL (photo-z calibration emission line sensitivities). In the first two cases, it interfaces to known galaxy properties to predict not just limiting magnitudes but also statistical properties (e.g. number counts or redshift distribution) of the detected and (in the case of WL) resolved sources.

The ETC is written in standard C and does not link to any external libraries. This should make it easier for users to install and run, as well as avoiding legal issues with copyrighted code.

The ETC also aspires to a high degree of realism, including:

- Full treatment of non-Gaussian galaxy and PSF profiles, including the effect of wavefront error.
- A model for the noise in NIR detectors, including propagation of both read and shot noise through the ramp fit, as well as a read noise floor.
- Inclusion of the major sources of backgrounds, including the zodiacal light, dark current, and telescope thermal emission.
- Integrating over the joint distribution of Hα line flux and $r_{\rm eff}$ when estimating the yield of the emission line survey; and incorporation of a galaxy catalog with a representative (magnitude, $z$, $r_{\rm eff}$) distribution for the WL source counts.
- Fisher matrix computation of galaxy ellipticity uncertainties.

Currently, however, computations are implemented in the "background-limited" regime. This is appropriate for galaxies near the limit of sensitivity, or even for the case near a SNR cut for WL. However for extreme values of parameters (e.g. requiring $\sigma_e < 0.02$ on a space mission) one may enter the "source-limited" regime. The code outputs basic data such as background levels and counts per exposure from objects of various magnitudes so that the user can diagnose this situation should it occur.

Suggestions for improvement, or additional effects that should be included, can be sent to the point of contact (Christopher Hirata).

This manual is deliberately verbose: it errs on the side of providing more information, and providing it repetitively. It also serves the dual role of reminding the author how everything works, hence §8 on data structures and functions. Most readers (who do not wish to write their own drivers) will need only the first several sections.

The user will note that this ETC does not live up to its title: one inputs an exposure time, and the code returns the observed flux limits and properties of galaxies (not the other way around). If you do not like the answer then you can increase the exposure time. The code was configured this way in part for simplicity, but also because many of the WFIRST programs prefer to cover sky at as fast a rate as possible while maintaining high data quality/redundancy and low overheads, or are constrained to the survey rate of another program.

## 2. Compiling the Code

The ETC is actually two calculators in one: an imaging/weak lensing calculator and a spectroscopy/redshift survey calculator. These calculators draw on a large number of similar functions (to compute zodiacal backgrounds, PSFs, etc.) and hence have a common source code, but are two separate executables distinguished by compiler flags.

The code should compile with the commands:

```
gcc exptimecalc.c -lm -Wall -O3 -o wletc.exe -DWL_MODE
gcc exptimecalc.c -lm -Wall -O3 -o baoetc.exe -DBAO_MODE
gcc exptimecalc.c –lm –Wall –O3 –o pzcaletc.exe -DPZCAL_MODE
```

which will generate the imaging ETC `wletc.exe` and the spectroscopy ETCs `baoetc.exe` and `pzcaletc.exe`.

### A. Additional flags

The code supports several additional flags that modify its behavior or allow additional user inputs. These can be appended on the compilation command after `–DWL_MODE` or `–DBAO_MODE` or `–DPZCAL_MODE` e.g.:

```
gcc exptimecalc.c -lm -Wall -O3 -o wletc.exe -DWL_MODE -DIN_DLON
```

The supported options are:

Options related to the SNR computations:

`–DUSE_SNRE`
Switches the BAO SNR computations from a cut on observed SNR to a cut on expected SNR. In v10 and later, the cut on SNRo is default and this option makes SNRe available. In v9 and earlier, all computations used a cut on SNRe.

Options related to the ecliptic longitude of observations:

-DIN_DLON

The default assumption is that observations are done at solar elongation ε = 90° (equivalently, the ecliptic longitude relative to the Sun is $\lambda-\lambda_\odot$ = 90°). This behavior can be changed with the -DIN_DLON flag, which allows the user to specify a different value of $\lambda-\lambda_\odot$.

-DTILT_BACK

This option replaces the default behavior with one that increases the ecliptic longitude relative to the Sun to the maximum legal value satisfying two constraints: (i) elongation ε ≤ 115°; and (ii) the angle between the curves of constant ecliptic latitude and direction to the Sun is ≤ 30° (thereby enabling opposite dispersion direction for configurations with a single spectrometer).

These two options should **not** be used together: in v7 and later an error message is generated.

Options to assume common values for inputs:

-DREAD5

This sets the detector characteristics to 5 electrons rms read noise floor and a dark current of 0.05 electrons per second per pixel, and does not prompt the user for these inputs.

-DWLCUT_DEFAULT

This sets the default WL galaxy catalog cuts: resolution factor > 0.4 and 1σ ellipticity error < 0.2, and does not prompt the user for these inputs.

Options to override assertions in the program:

-DWFE_OVERRIDE

Overrides the limit on RMSWFE (<0.075$\lambda_{min}$) in WL mode. This invalidates the computation of shape measurement results, and hence turns off these outputs, but is still useful for calculating sensitivities for the imaging survey. Has no effect on BAO mode.

Options to send additional data to a file:

-DOUT_WL_CAT

Sends the weak lensing source catalog to an output file.

Options to change the noise model:

-DNO_OBJ_CONTINUUM

Turns off the source continuum contribution to the noise model (usually a minor effect, of order ~1% since in wide-bandwidth slitless spectroscopy the galaxy continuum is almost always far below the sky background).

`-DLOGICAL_READ_FLOOR`
   This option changes the default behavior of the read noise floor (added in quadrature to the theoretical Poisson + read noise error) to one where it replaces the read noise if the latter falls below floor. See §7.F for details.

Options related to optics:

`-DWFE_HOPKINS_EXP`
   Switches the aberration model from default (conservative) to the exponential Hopkins distribution.

Options related to the user interface:

`-DKEEP_WINDOW_OPEN`
   This option requires the user to hit the *<return>* key in order to exit at the end of the program. This is useful in environments where the output window closes when the program exits.

Options no longer supported:

The `-DCCD_MODE` option no longer exists (removed in v9), since the detector type is specified at runtime.

# 3. Running the Code

Both the WL and BAO ETCs have a similar set of inputs, with the exception of the galaxy densities. Input units are specified at the prompt where appropriate.
The code prints in the first line which compiler option(s) are used.

### A. Specifying the telescope configuration

The first input will be the telescope configuration. There are two options here: one may either select a "generic" configuration, with the primary mirror diameter, pixel scale, and throughput specified by the user; or a "from file" configuration, where this information is read from a data file. The "generic" configuration is useful for simple comparisons, but the "from file" option has greater flexibility (including support for realistic wavelength-dependent throughputs) and is therefore the better choice for detailed design work.
A "generic" configuration input might look like:

```
./baoetc.exe
Enter telescope configuration [0=generic, 1=from file]: 0
Enter aperture outer diameter (meters): 1.5
Enter central obscuration (fractional - linear): 0.5
Enter pixel scale (arcsec): 0.37
Enter throughput (total, all orders): 0.6
```

```
Enter throughput (1st order): 0.6
```

The aperture outer diameter *D* should be that of the limiting aperture. In cases with an outer pupil mask, *D* should be appropriately reduced (i.e. it would be less than the physical outer diameter of the primary mirror). The central obscuration υ is in linear units (i.e. the inner diameter is υ*D*); one may set υ = 0 for an unobstructed system. The pixel scale *P* is in arc seconds.

The throughput numbers take two forms: the total throughput, and (for the BAO ETC) the 1st order throughput. The $n^{\text{th}}$ order throughput $f_n$ is defined as the ratio of the effective area $A_{\text{eff},n}$ in the $n^{\text{th}}$ order to the geometric area of the annulus.

$$A_{\text{eff},n} = f_n \frac{\pi}{4}(1-\upsilon^2)D^2.$$

For the WL ETC, the throughput controls both signal and background, whereas for the BAO ETC the all-order throughput $f_{\text{all}}$ controls the background while only the 1st order throughput $f_1$ is used to compute the signal. For prism spectroscopy one normally has $f_1 = f_{\text{all}}$ while a grism would have $f_1 < f_{\text{all}}$ (but see the note below on scattered light).

A "from file" input might look like:

```
./baoetc.exe
Enter telescope configuration [0=generic, 1=from file]: 1
Input file name: data/JDEM_Omega_nomask_SpC.dat
Using configuration: JDEM_Omega_no_pupil_mask_110323
```

The input file contains the geometrical parameters (*D*, υ, and *P*) as well as tables of the total throughput $f_{\text{all}}$ and the desired-order throughput ($f_1$ for BAO and $f_0$ for WL). The file format is described in §3.B. It is also possible to incorporate thermal parameters in the configuration file.

Several points are in order regarding the throughputs.

- The current version of the ETC does not include spiders. However, it is possible to approximately include a spider with appropriate adjustments to the inputs. A spider that blocks a fraction $f_{\text{sp}}$ of the annulus has two major effects: one is to reduce the effective area by a factor of $1-f_{\text{sp}}$; and the other is to scatter a fraction $f_{\text{sp}}$ of the remaining light into diffraction spikes (which are not useful for detecting objects or emission lines). We thus recommend that configurations with spiders incorporate both of these effects in the throughput curve: $f_{\text{all}}(\lambda)$ is reduced by a factor of $1-f_{\text{sp}}$ relative to what it would be if the spider were magically removed; and $f_0(\lambda)$ or $f_1(\lambda)$ is reduced by a factor of $(1-f_{\text{sp}})^2$. A consequence of this is that for an imager we have $f_0(\lambda)/f_{\text{all}}(\lambda) = 1-f_{\text{sp}}$ instead of 1.
- The ETC does not currently include the production of multiple charges per photon, which may be significant at very blue wavelengths. This should not be significant in the range of wavelengths recommended for the WFIRST WL and BAO programs, λ > 1 μm.

If the `WFE_INPUT` option is used (it is on by default in the code in v6 and later using `#define WFE_INPUT`, and can be turned off only by removing this line), then the code prompts the user for the rms wave front error in μm. The effect of an aberration is not uniquely determined by the rms wave front error, so the code uses a conservative fitting function assuming the error is in low-order (≤3) Zernike modes (see §7.I for details).

### B. Telescope configuration file format

The telescope configuration file contains basic properties such as the aperture and pixel scale, as well as a table of throughput as a function of wavelength. It is also possible to incorporate thermal properties of the telescope in the configuration file that override the default settings of the ETC. <u>Configuration files are always limited to lines of no more than 254 characters.</u>

An explicit example of a configuration file is `data/JDEM_Omega_nomask_SpC.dat`, which has the form[1]:

```
# JDEM Omega, no pupil mask
# based on 'Throughput table JDEM Omega 110322_cj.xlsx'
# Received: 2011/03/23
#
# This has no pupil mask, so we have 1.5 m outer diameter and 50%
# linear obscuration. The spider is included in the throughput (all
# orders) numbers. For the '1st order' throughput I simply squared
# these (i.e. the spider blocks 2.45% of the radiation and scatters
# an additional 2.45% into the far diffraction wings where it is
# useless but still contributes to background).
#
JDEM_Omega_no_pupil_mask_110323
1.5  0.5  0.37  30
1.1000   0.3947   0.3850
1.1310   0.3973   0.3876
1.1621   0.4371   0.4263
…
2.0000   0.5618   0.5481
```

The file begins with a set of comment lines that start with # (note that under the current code structure, all comment lines must be at the beginning of the file: they may not be interspersed with the input data). The first non-comment line is a configuration label (in this case "`JDEM_Omega_no_pupil_mask_110323`") that must begin with a non-special character (alphanumeric or underscore). The next line has four numbers: in order these are the outer diameter $D$ (in meters); the central obscuration $\upsilon$; the pixel scale $P$ (in arcsec); and the number of grid points $N$ at which the throughput curve is defined. Following this is a table with $N$ lines each with three numbers: the wavelength $\lambda$ in μm; the total throughput $f_{\mathrm{all}}(\lambda)$; and the desired-order throughput $f_0(\lambda)$ or $f_1(\lambda)$ (for imagers or spectrometers, respectively).

---

[1] The data for this input file was provided by Dave Content and Cliff Jackson.

The throughput table is interpolated linearly between grid points. It is assumed constant outside the range of the table if the user requests such a wavelength, but of course it is not recommended to use the ETC outside the range of the throughput table unless this represents the expected behavior of the system!

Non-default thermal emission parameters can be specified in the configuration file with a `!THERMAL` line. This line must appear after the end of the comments and before the configuration label. An example would be:

```
# JDEM Omega, no pupil mask
# This configuration has thermal parameters.
#
!THERMAL 219. 0.02 1.10 0 219. 0.08 5e-3
JDEM_Omega_no_pupil_mask_110323 (ImC)
1.5 0.5 0.18 30
0.4000   0.4362   0.4255
0.4552   0.5203   0.5076
0.5103   0.5562   0.5426
…
2.0000   0.7109   0.6934
```

The `!THERMAL` line, like all lines in the configuration file, is limited to no more than 254 characters. The 7 parameters specified are, in order:

- The telescope temperature, $T_{tel}$ (in Kelvin);
- The emissivities per surface ε of the primary and secondary mirrors;
- The ratio ρ of outer diameter of the illuminated region to that of the primary mirror;
- An integer flag indicating whether there is a cold pupil mask (1 = yes, 0 = no);
- The temperature of the aft optics upstream of the filter, $T_{aft}$ (in Kelvin);
- The summed emissivity $ε_{aft}$ of the aft optics; and
- The count rate from the filter and downstream components $I_{down}$ (in $e^-$/pix/s).

The `!THERMAL` line overrides the default settings of all these parameters. See §7.G for a detailed description of how these are used.

### C. Specifying the observing strategy and detector characteristics

Having described the telescope, one must next specify the properties of the filter, detector, and observational circumstances. The following set of inputs are requested (here with sample inputs):

```
Enter detector type [e.g. 0=H2RG]: 0
Enter pointing jitter (arcsec rms per axis): 0.04
Enter minimum wavelength (microns): 1.1
Enter maximum wavelength (microns): 2.0
Enter single exposure time (s): 150
Enter read noise floor (effective e- rms per pixel): 5
Enter dark current (e-/pix/sec): 0.05
Enter ecliptic latitude (degrees): 45
Enter Galactic reddening, E(B-V) (magnitudes): 0.05
```

```
Enter number of exposures: 6
```

The currently available detector types (and the index for inputting them) are:

| Index | Type | Pixel Scale [µm] | Charge Diffusion [µm rms/axis] | NIR Detector Parameters |
|---|---|---|---|---|
| 0 | H2RG (NIR) [2,3] 32 channels | 18.0 | 2.94 | $t_f$ = 1.3 s; $\sigma^2_{read}$ = 200 |
| 1 | e2v CCD | 12.0 | 5.00 | N/A |
| 2 | H4RG (NIR) [4] 32 channels | 10.0 | 2.94 | $t_f$ = 5.24 s; $\sigma^2_{read}$ = 200 |
| 3 | H4RG (NIR) 64 channels | 10.0 | 2.94 | $t_f$ = 2.62 s; $\sigma^2_{read}$ = 200 |

The overall PSF is taken to be the convolution of the diffraction pattern from the annulus, aberrations if nonzero wave front error is specified, the detector response (pixel tophat plus charge diffusion), and the pointing jitter. The detector is assumed to be an H2RG with 18 µm pixels (see §7.I for details).

The wavelength range ($\lambda_{min}$, $\lambda_{max}$) and the time per single exposure ($t_1$) are also specified. In WL mode, the code will also prompt the user for an imaging filter throughput (you can enter 1 if the filter throughput is already included in the telescope configuration file; this will give a warning message, which you can ignore if you intend for no additional filter losses to be considered).

The read noise floor is in effective electrons rms per exposure. For CCDs this is the entire read noise; for NIR detectors this is added in quadrature to the statistical error from ramp fitting. In principle this is determined by the readout mode and the exposure time, but at present this and the dark current are allowed to be user inputs.

The position of the observation on the sky is required in order to calculate the zodiacal foreground radiation. This in principle requires the ecliptic latitude β and the difference in ecliptic longitudes of the observation and the Sun, $\lambda-\lambda_\odot$. In default mode the latter is set to 90°, but can be independently specified if the –DIN_DLON compilation flag is used.

---

[2] The charge diffusion length is measured to be 1.87 µm by Barron *et al.* (2007). The standard deviation of the sech profile is π/2 times the diffusion length. Using the sech profile instead of a Gaussian would increase the MTF by only 0.2% at 1 cycle/arcsec and 2% at 2 cycles/arcsec at the 0.18 arcsec plate scale.

[3] The additional NIR detector parameters assume a read noise of 20 $e^-$ per CDS and a readout rate of 100 kHz for each of 32 channels.

[4] The H4RG is a new detector and is not yet at TRL 6. It is therefore not baselined for WFIRST. The parameters above are considered reasonable targets for successful H4RG development program. They assume that the charge diffusion remains fixed in physical units (µm) and that the read noise per CDS remains fixed at 20 $e^-$ per CDS (although this is not yet achieved). The readout rate is kept at 100 kHz, and either 32 or 64 channels are used (the latter is an open trade for the H4RG options).

The Galactic dust column is specified by the color excess $E(B-V)$. All output sensitivities are quoted in terms of extinction-corrected (rather than observed) fluxes and magnitudes. If observed flux limits are desired (e.g. for Galactic objects that may be in front of some or all of the dust column, or for consistency with other surveys that usually quote sensitivities in observed fluxes) then one should input $E(B-V) = 0$.

The number of exposures is $N_{exp}$ (6 in the above example); the total "live" time is $N_{exp}t_1$.

### D. <u>**WL-specific inputs**</u>

The WL ETC requires three inputs from the user: a cut $R_{min}$ on the resolution factor $R$ of the galaxies; a cut $\sigma_{e,max}$ on the ellipticity error $\sigma_e$ of galaxies to be measured; and the galaxy catalog to use to estimate source densities. Example inputs are:

```
Enter minimum resolution factor: 0.4
Enter maximum ellipticity error: 0.2
Enter input galaxy catalog file: data/CMC1.dat
```

The resolution factor $R$ is defined on a scale of 0 (galaxy small compared to the PSF) to 1 (galaxy large compared to the PSF); the definition used here is

$$R = \left(1 + \frac{EE50_{PSF}^2}{r_{eff,gal}^2}\right)^{-1}.$$

For Gaussian profiles, this definition agrees exactly with Bernstein & Jarvis (2002). Galaxies with $R << 1$ usually have unreliable shapes. In the SDSS lensing analyses (see e.g. Mandelbaum *et al.* 2005) the standard cut on $R$ was ⅓; the recent WFIRST forecasts have used 0.4. The ellipticity is defined using the convention

$$(e_1, e_2) = \frac{a^2 - b^2}{a^2 + b^2}(\cos 2\chi, \sin 2\chi),$$

where $a$ and $b$ are the semi-major and semi-minor axes of the galaxy and $\chi$ is the position angle of the major axis. Ellipticity errors are defined as rms per component.

Even if a galaxy is well-resolved, it will not be usable for WL if its SNR is too low to recover a shape. For this reason, we include a maximum ellipticity error $\sigma_{e,max}$. The computation of the ellipticity uncertainty assumes an exponential profile galaxy and does a full integration over scales as described in the methodology section. It does <u>not</u> yet account for undersampling considerations. The specifics of whether a given observing strategy can recover full sampling while preserving the original PSF (or smoothing it in a controlled way) must be evaluated using an image reconstruction simulator (e.g. Rowe *et al.* 2011). Finally, we impose a minimum SNR for the galaxy: in the current version of the code this is hard-wired at SNR > 18.[5]

---

[5] The cut is set by the `#define WL_SNR_MIN 18.0` macro.

The input galaxy catalog is a file. An example is the file `CMC1.dat`:

```
# Cosmos Mock Catalog (Jouvel et al 2009), 435W-606W-775W-850LP-110-160-Ks
7 438972 1.24
0.431177 0.587964 0.768801 0.901067 1.103397 1.601053 2.160917
     1 0.96 0.1463 27.50 26.80 26.11 25.80 25.44 25.19 24.89
     9 3.48 0.1283 27.72 26.22 25.94 25.94 25.84 25.22 24.33
    10 0.48 0.2228 26.82 26.03 25.45 25.20 24.98 24.63 24.43
…
537990 2.20 0.1404 27.58 26.96 26.52 26.07 25.34 24.27 23.39
```

The file begins with a set of comment lines (starting with #); comment lines must be at the beginning, and not interspersed through the file. The first non-comment line has three entries: a number $N_\lambda$ of wavelength nodes at which galaxy SEDs are specified; a number of galaxies in the catalog; and the catalog area in deg². Next comes a list of the node wavelengths in microns. Each subsequent line has $3+N_\lambda$ entries; in order, these are:

- An identification number (an integer). It is used only if an output file is specified (using the `-DOUT_WL_CAT` option). Its purpose is to allow external programs to forecast the overlap of the lensing source catalog with other catalogs (e.g. successful LSST+WFIRST photometric redshifts).
- The redshift $z$.
- The effective radius $r_{eff}$ in arcsec.
- The AB magnitude at each of the wavelength nodes.

The code constructs an SED by connecting the nodes with power laws. For nodes spaced by typical filter widths, this provides a simple and fast way of estimating galaxy yields and $dN/dz$ in a WFIRST-like filter. An error message is returned if the specified bandpass is not contained within the nodes in the galaxy catalog file.

The galaxy catalog `CMC1.dat` is used for most analyses is extracted from the CMC (Jouvel *et al.* 2009; taking $r_{eff}>0$ and $z>0$ objects). It is based on the (non-randomized) magnitudes for a suite of 7 filters ranging from 0.43—2.16 μm.

If the `-DOUT_WL_CAT` option is specified, the code will ask for an output file name. The output file will have the form:

```
    13 0.1200 0.2027 0.17099
    41 0.1600 0.2399 0.12337
   330 0.0800 0.4840 0.10593
   576 0.1600 1.4943 0.06834
…
 537695 2.8000 0.2092 0.17434
```

The columns are the galaxy ID number; the redshift, $r_{eff}$ in arcsec; and $\sigma_e$.

### E. BAO-specific inputs

The BAO ETC further requests the minimum SNR for an Hα line detection and a choice of galaxy population model:

```
Enter significance cut (number of sigmas): 7.0
Enter galaxy population model [suggested = 32]: 0
Using luminosity function model: Geach et al (2009), 50% confidence
```

The overall galaxy population is described by a joint distribution of Hα line luminosity $L_{H\alpha}$ and effective radius $r_{eff}$. Equivalently, this can be specified by a combination of the HαLF and the conditional size probability distribution $P(r_{eff}|L_{H\alpha},z)$. The code is currently configured to take a population model that is a non-negative integer of the form $10i+j$, where $i$ indicates the choice of HαLF and the single-digit $j$ indicates the choice of conditional size probability distribution. We currently recommend using the most up-to-date luminosity function with all corrections, which is model 32.

In v9 and later, the redshift survey completeness[6] (as a fraction of the number of emission lines theoretically detected above a certain number of σ) is also requested:

```
Enter completeness: 0.5
```

The recommended (current) size distribution based on CMC is $j = 2$. In v7 and earlier (including forecasts for the Green *et al.* 2011 WFIRST SDT Interim Report), a size distribution was used based on the original version of the CMC. This older model is still available by setting $j = 0$.

Hα luminosity functions:

The majority of the luminosity functions are in the Schechter form. This form supposes that the probability of finding a galaxy per unit comoving volume per unit $L_{H\alpha}$ is

$$dP = \phi_* \frac{L^\alpha}{L_*^{\alpha+1}} e^{-L/L_*} \, dV \, dL,$$

where the faint-end slope α, the characteristic density $\phi_*$, and the characteristic luminosity $L_*$ are parameters.

The currently available HαLFs are:

- $i = 0$: The reference model for the Hα luminosity function is that of Geach *et al* (2010). It has a Schechter form with parameters $\phi_* = 1.35 \times 10^{-3}$ Mpc$^{-3}$, $\alpha = -1.35$, and $L_* = 5.1 \times 10^{34} (1+z)^{3.1}$ W (at $z < 1.3$) or $6.8 \times 10^{35}$ W ($z > 1.3$).
- $i = 1$: This is the one-sided 90% confidence lower limit of the Geach *et al.* (2010) HαLF. The only change to the parameters is that $L_*$ has been reduced to $5.1 \times 10^{34}(1+z)^{2.59}$ W ($z \leq 1.3$) and is constant at $z \geq 1.3$.
- $i = 2$: This is the Geach *et al.* (2010) HαLF with $\phi_*$ reduced by a factor of 1.257 from its nominal value. It is a more conservative estimate than the Geach *et al.* (2010) central value, but over most of the redshift range it is not as conservative as $i = 1$.[7,8]

---

[6] The behavior of v8 and earlier was that the completeness was implicitly set to 1, and it was the responsibility of the user to multiply the output by the expected completeness.
[7] Y. Wang, private communication.

- *i* = 3: The Sobral *et al.* (2012) HαLF based on ground-based narrow band searches. The Schechter function parameters have had the following corrections applied: [i] correction to the WMAP 5-year cosmology (Sobral *et al* used $H_0$ = 70 km s$^{-1}$ Mpc$^{-1}$ and $\Omega_m$ = 0.3); [ii] conversion back to observed (rather than extinction-corrected) Hα luminosities; and [iii] aperture-corrected to integrated luminosities rather than 2 or 3 arcsec aperture luminosities in 0.8 arcsec FWHM seeing (Kolmogorov profile PSF, and exponential profile galaxies using the median size at that redshift at $F_{H\alpha}$ = 10$^{-19}$ W m$^{-2}$ and the *j* = 2 size model). The resulting luminosity function parameters are shown in Table 1. The parameters log$_{10}$ ϕ$_*$ and log$_{10}$ $L_*$ are linearly interpolated in *z* (they are extrapolated at *z* < 0.40, but this is not relevant to the wavelength ranges to be observed by Euclid or WFIRST). The HαLF is assumed constant at *z* > 2.23 (this is conservative: we do not know whether $L_*$ continues to rise at higher *z*) but this only affects the highest-redshift WFIRST bins.

**Table 1: The Schechter function parameters of the Sobral *et al*. (2012) HαLF (model *i* = 3), and the corrections applied.**

|  | *z* = 0.40 | *z* = 0.84 | *z* = 1.47 | *z* = 2.23 |
|---|---|---|---|---|
| Faint end slope α | −1.60 | −1.60 | −1.60 | −1.60 |
| Original log$_{10}$ ϕ$_*$ [Mpc$^{-3}$] | −3.12 | −2.47 | −2.61 | −2.74 |
| Original log$_{10}$ $L_*$ [erg s$^{-1}$] | 42.09 | 42.25 | 42.56 | 42.86 |
| Cosmological volume correction to ϕ$_*$ [dex] | +0.01 | −0.01 | −0.02 | −0.03 |
| Cosmological distance correction to $L_*$ [dex] | −0.01 | 0.00 | +0.01 | +0.01 |
| Undoing extinction correction to $L_*$ [dex] | −0.40 | −0.40 | −0.40 | −0.40 |
| Aperture correction to $L_*$ [dex] | +0.02 | +0.07 | +0.07 | +0.06 |
| Final log$_{10}$ ϕ$_*$ [Mpc$^{-3}$] | −3.11 | −2.48 | −2.63 | −2.76 |
| Final log$_{10}$ $L_*$ [erg s$^{-1}$] | 41.71 | 41.92 | 42.23 | 42.54 |

The code prints a description of the chosen HαLF to the screen so that the user can verify the correct selection.

Conditional size distributions:

The **recommended** *j* = 2 size distribution is based on the CMC (Jouvel *et al*. 2009), updated as of August 15, 2011 (Zoubian *et al*., in preparation). The objects with redshifted Hα wavelength between 1.1—2.0 μm were selected, and split into bins based on the Hα flux $F_{H\alpha}$ [W m$^{-2}$]. For five bins in log$_{10}$ ($F_{H\alpha}$ [W m$^{-2}$]) centered at −19.0(0.2)−18.2, and in two wavelength ranges (1.1—1.5 and 1.5—2.0 μm) the effective radii are well fit by

$$\log_{10} \frac{r_{\rm eff}}{\rm arcsec} = -0.66 + 0.06\xi + (0.27 + 0.01\xi)\left(19 + \log_{10} \frac{F_{H\alpha}}{\rm W\,m^{-2}}\right) \pm 0.2,$$

where

---

[8] This was the model used for the WFIRST IDRM studies as well as the 2011 Euclid studies.

$$\xi = \frac{1.75 - \lambda_{H\alpha,obs}[\mu m]}{0.45},$$

and where 0.2 dex denotes the 1σ dispersion. Note that the effective radii were measured in the F814W band (rest frame ~*u* band continuum), whereas for WFIRST or Euclid it is actually the effective radius in the Hα line (also a tracer of star formation) that is relevant. Since we do not have sufficient data on the latter, the CMC radii are probably the most accurate we have at this time, and we adopt the above lognormal $P(r_{eff}|F_{H\alpha})$.

The $j = 0$ size distribution is based on the original version of the CMC and had a wavelength range of 1.5—2.0 µm, but the construction procedure was otherwise equivalent. This distribution is

$$\log_{10}\frac{r_{eff}}{arcsec} = -0.62 + 0.25\left(19 + \log_{10}\frac{F_{H\alpha}}{W\ m^{-2}}\right) \pm 0.2.$$

### F. **PZCAL-specific inputs**

There are no inputs unique to the PZCAL mode. In general the inputs are identical to the BAO mode, *except* that the galaxy population model is not required.

## 4. Code Outputs: WL ETC

The WL ETC returns three blocks of information:

- The general properties of the imaging survey.
- The magnitude and size thresholds for shape measurement.
- The demographics of the shape galaxy population.

The properties of the imaging survey are returned in the first block:

```
General properties:
PSF EE50:                    2.96232E-01 arcsec
Spatial frequency cutoff:    8.44514E-01 cycles/pix
Sampling case:               Weakly undersampled
Min usable galaxy r_eff:     2.41872E-01 arcsec
Sky background flux:         3.03466E-01 e-/pix/s
Thermal background flux:     3.37757E-02 e-/pix/s
  telescope:                 2.31562E-02 e-/pix/s
  upstream:                  5.61950E-03 e-/pix/s
  downstream:                5.00000E-03 e-/pix/s
Noise variance per unit solid angle in one exposure:
            sky only:        1.68592E+03 e-^2/arcsec^2
               total:        3.90252E+03 e-^2/arcsec^2
Source counts per exposure:
         at AB mag 20:       2.11038E+04 e-
         at AB mag 21:       8.40157E+03 e-
         at AB mag 22:       3.34473E+03 e-
```

```
         at AB mag 23:    1.33156E+03 e-
         at AB mag 24:    5.30104E+02 e-
         at AB mag 25:    2.11038E+02 e-
5 sigma pt src threshold:    25.694 mag AB
```

The most useful measure of the resolution of the survey for sources of moderate SNR (i.e. such that one would not attempt deconvolution) is EE50. The sampling condition is given by the spatial frequency cutoff, which the WL ETC returns in units of cycles per pixel; recall that "oversampling" at the native pixel scale requires this cutoff to be <0.5. The ETC describes undersampled systems as either "weakly undersampled" or "strongly undersampled" depending on whether there are unaliased Fourier modes in the image or not. The algorithms required for data processing branch into the three different cases – in the latter two cases multiple defect-free dithers become necessary to measure the shapes. However, **the ETC does not penalize surveys with inadequate sampling. It is the user's responsibility to independently check that their chosen dithering pattern will be successful**.

The minimum usable galaxy size according to the user-specified cut on resolution factor is given, as are the sky backgrounds and contributions to noise variance. Many imaging cases that we have looked at are actually limited by detector (dark + read) noise rather than sky; the outputs are designed to make clear which regime one is in. The number of counts from sources at different magnitudes is specified, as is the often-quoted 5σ point source sensitivity limit (this is not directly relevant for WL, which requires >>5σ and an extended source, but is a useful point of comparison).

Next comes a table of galaxy properties as a function of the effective radius:

| r_eff | Om_eff | penalty factor | resolution factor | lim mag (shapes) | S/N at shape lim mag |
|--------|--------|----------------|-------------------|------------------|----------------------|
| arcsec | arcsec2 | | | AB | |
| 0.25119 | 1.7427 | 5.2704 | 0.41827 | 23.49161 | 22.957 |
| 0.28184 | 1.9602 | 4.6476 | 0.47512 | 23.49603 | 21.558 |
| 0.31623 | 2.2241 | 4.1279 | 0.53261 | 23.49184 | 20.317 |
| 0.35481 | 2.5453 | 3.6967 | 0.58926 | 23.47848 | 19.227 |
| 0.39811 | 2.9375 | 3.3402 | 0.64363 | 23.45574 | 18.276 |
| 0.44668 | 3.4175 | 3.0460 | 0.69454 | 23.39012 | 18.000 |
| 0.50119 | 4.0062 | 2.8030 | 0.74110 | 23.30384 | 18.000 |
| 0.56234 | 4.7299 | 2.6021 | 0.78278 | 23.21368 | 18.000 |
| 0.63096 | 5.6214 | 2.4354 | 0.81939 | 23.11995 | 18.000 |
| 0.70795 | 6.7216 | 2.2967 | 0.85100 | 23.02291 | 18.000 |
| 0.79433 | 8.0819 | 2.1808 | 0.87790 | 22.92286 | 18.000 |
| 0.89125 | 9.7665 | 2.0836 | 0.90052 | 22.82008 | 18.000 |
| 1.00000 | 11.8559 | 2.0017 | 0.91933 | 22.71483 | 18.000 |

This provides, as a function of $r_{\rm eff}$, the effective solid angle of the galaxy after PSF convolution $\Omega_{\rm eff}$; the shape penalty factor $f_{\rm pen}$, defined by

$$\sigma_e = \frac{2\sqrt{f_{\rm pen}}}{\rm SNR},$$

which is 1 for a well-resolved Gaussian galaxy (Bernstein & Jarvis 2002, Eq. 3.15) and approaches ∞ for unresolved objects; the resolution factor $R$; the limiting AB magnitude for shape measurements; and the SNR for objects at the limiting magnitude for shapes. The penalty factor is computed from the Fisher matrix for the galaxy ellipticity, as described in the methodology section.

Finally, using the galaxy catalog provided one may derive the population of detected galaxies as a function of redshift. Their redshift distribution (currently in bins[9] of width $\Delta z$ = 0.2, up to a maximum of $z$ = 3) and total surface density are returned. There are two columns, distinguished by $N$ versus $N_{\rm eff}$. Here $N$ is the total number of galaxies with measured shapes; $N_{\rm eff}$ is defined by down-weighting galaxies with non-negligible shape measurement noise, i.e.

$$N_{\rm eff} = \sum_{\rm galaxies} \frac{0.4^2}{0.4^2 + \sigma_e^2},$$

where 0.4 is the assumed intrinsic ellipticity dispersion.

```
|   z   |  dN/dz/dA  |dNeff/dz/dA|
|       |   deg^-2   |   deg^-2  |
 0.100   1.23105E+04  1.13740E+04
 0.300   4.34395E+04  4.05262E+04
 0.500   3.51008E+04  3.26792E+04
 0.700   5.21573E+04  4.82890E+04
 0.900   5.27903E+04  4.88995E+04
 1.100   3.64879E+04  3.34505E+04
 1.300   2.95040E+04  2.67773E+04
 1.500   2.44677E+04  2.21167E+04
 1.700   1.28468E+04  1.14838E+04
 1.900   7.81855E+03  6.96697E+03
 2.100   1.03629E+04  9.09031E+03
 2.300   4.06048E+03  3.55085E+03
 2.500   4.10887E+03  3.59047E+03
 2.700   2.42339E+03  2.10688E+03
 2.900   1.87903E+03  1.62101E+03

Weak lensing n:                 6.59516E+04 gal/deg^2
                        =       1.83199E+01 gal/arcmin^2
Weak lensing n_eff:             6.05045E+04 gal/deg^2
                        =       1.68068E+01 gal/arcmin^2
```

## 5. Code Outputs: BAO ETC

---

[9] The bin sizes are currently set to 0.199999, so that input catalogs such as the CMC that have quantized galaxy redshifts do not give large errors from galaxies that fall into 0 or 2 of the bins.

The BAO ETC returns a table of galaxy yields as well as some summary information. An example yields table is:

```
    z  |lambda| EE50  |  dV/dz/dA  | Flim@0.30"| n targets |   dN/dz/dA |geom mean F| siglnF |skew lnF|kurt lnF|Num ph 1exp|
       |  um  |arcsec| Mpc3/deg2  |    W/m2   |   Mpc^-3  |    deg^-2  |    W/m2   |        |        |        | 1e-19 W/m2|
 1.000 1.3130 0.3005 8.29240E+06 3.22901E-19 3.79620E-04 3.14796E+03 6.36092E-19   0.5068   0.7603   0.2992 8.26072E+01
 1.050 1.3458 0.2966 8.67969E+06 3.05516E-19 3.92039E-04 3.40278E+03 5.96399E-19   0.5105   0.7642   0.2997 8.72193E+01
 1.100 1.3787 0.2924 9.04621E+06 2.77963E-19 4.29332E-04 3.88383E+03 5.41519E-19   0.5223   0.7697   0.2919 9.57537E+01
 ...
 2.650 2.3962 0.2917 1.29985E+07 1.46194E-19 9.35757E-05 1.21635E+03 1.74560E-19   0.3438   0.9389   0.7847 1.90322E+02
```

The table gives the properties of the available galaxy sample as a function of redshift $z$, over the entire range for which Hα is available. The columns are:

i. The redshift $z$.
ii. The wavelength λ (in μm) at which Hα is observed.
iii. The EE50 radius of the PSF (in arcsec).
iv. The conversion from area to comoving volume, $dV/dz/dA$ (in Mpc$^3$ deg$^{-2}$).
v. The MDLF for a source with $r_{\rm eff}$ = 0.3 arcsec (in W m$^{-2}$).[10]
vi. The comoving density of detectable galaxies $n$ (in Mpc$^{-3}$).
vii. The area density of detectable galaxies per unit solid angle per Δ$z$ (in deg$^{-2}$).
viii. The geometric mean Hα flux $F_{\rm Hα}$ of the detectable galaxies (in W m$^{-2}$).
ix. The standard deviation of ln $F_{\rm Hα}$ of these galaxies.
x. The skewness of ln $F_{\rm Hα}$ of these galaxies.
xi. The kurtosis of ln $F_{\rm Hα}$ of these galaxies.
xii. The number of collected electrons per exposure for a 10$^{-19}$ W m$^{-2}$ emission line.

It is important to note that galaxies can be described as "detectable," which does not equate to "detected" – one must still multiply by the completeness $e$, which is generally <1 due to e.g. collisions with star traces or image defects. Also the MLDF is provided simply for reference: while an effective radius of 0.3 arcsec is typical it is not universal, and the code integrates the number of galaxies within the ($F_{\rm Hα}$,$r_{\rm eff}$) plane.

The code also returns summary statistics:

```
Summary statistics:
Sky background flux:       3.08083E+00 e-/pix/s
Thermal background flux:   1.38794E-01 e-/pix/s
  telescope:               8.75479E-02 e-/pix/s
  upstream:                2.12459E-02 e-/pix/s
  downstream:              3.00000E-02 e-/pix/s
Noise variance per unit solid angle in one exposure:
           sky only:       3.37564E+03 e-^2/arcsec^2
              total:       4.65012E+03 e-^2/arcsec^2
Available galaxy density:  4.77475E+03 gal/deg^2
```

These include useful ancillary parameters such as the sky count rate, the noise variance per single exposure (including just the sky, and including dark current + read noise; the comparison is useful for determining whether one is sky-limited, which is only the case for some of the viable range of NIR space mission parameters), and the total density of detectable galaxies.

---

[10] Note that 1 W m$^{-2}$ = 1000 erg cm$^{-2}$ s$^{-1}$.

# 6.  Code Outputs: PZCAL ETC

In PZCAL mode, the ETC assumes that it is searching for emission lines and simply reports a table of line sensitivities across the spectral bandpass. The table is presented in 28 steps in $\lambda$[11] and $r_{eff}$ from 0.0 (point source) to 1.0 arcsec in steps of 0.1 arcsec.

A typical data table and associated summary statistics is:

```
Limiting fluxes in W/m2 vs wavelength and galaxy size
lambda|    0.0"   |    0.1"   |    0.2"   |…|    1.0"   |
  um  |           |           |           |…|           |
0.6000 1.71123E-19 2.15580E-19 2.97796E-19 … 1.08921E-18
0.6500 1.56623E-19 1.97578E-19 2.72938E-19 … 9.95079E-19
0.7000 1.34458E-19 1.69790E-19 2.34528E-19 … 8.52248E-19
0.7500 1.16880E-19 1.47663E-19 2.03902E-19 … 7.38483E-19
0.8000 1.04460E-19 1.32026E-19 1.82229E-19 … 6.57739E-19
…
2.0000 4.95831E-20 5.81933E-20 7.53537E-20 … 2.43001E-19

Summary statistics:
Sky background flux:       1.11131E+00 e-/pix/s
Thermal background flux:   6.10119E-02 e-/pix/s
  telescope:               1.79270E-02 e-/pix/s
  upstream:                3.80850E-02 e-/pix/s
  downstream:              5.00000E-03 e-/pix/s
Noise variance per unit solid angle in one exposure:
            sky only:      1.02899E+04 e-^2/arcsec^2
               total:      1.47232E+04 e-^2/arcsec^2
```

The summary statistics table is identical to that of BAO mode except for the galaxy yields (which are not computed in PZCAL mode).

# 7.  Detailed Methodology and Assumptions

### A. <u>Background cosmology</u>

The background cosmology assumed for number density yields (in the BAO mode) is the same as that of the FoMSWG (Albrecht *et al.* 2009; chosen for consistency), itself derived from the WMAP 5-year parameter constraints (Dunkley *et al.* 2009). The parameters are $H_0$ = 71.903 km/s/Mpc, $\Omega_m$ = 0.25648, and zero spatial curvature.

The WL mode uses an input galaxy catalog and does not currently utilize the cosmological parameters.

---

[11] This can be changed by modifying the `nstep = 28;` statement in the main program.

### B. Sky brightness

The sky brightness is taken to be that of the zodiacal light, which is taken to be a function not just of wavelength but also ecliptic latitude $\beta$ and longitude relative to the Sun $\lambda-\lambda_\odot$. The brightness $I_\lambda$ at $\lambda = 0.5$ μm was interpolated from Table 17 of Leinert *et al*. (1998). The sky brightness at redder wavelengths was determined using the solar spectrum (Colina *et al*. 1996) and the correction for the color of the zodiacal light (relative to the solar spectrum), Eq. (22) of Leinert *et al*. (1998). A similar method was used in the SNAP model (Aldering 2001).

The internal routines in the ETC have the capability to run at any ecliptic longitude, but by default the driver assumes an elongation of $\varepsilon = 90°$ (implying $\lambda-\lambda_\odot = 90°$). The zodiacal light is thus a function of wavelength and ecliptic latitude. Other variations discussed in Leinert *et al* are neglected (e.g. annual variation, discrete structures).

### C. Galactic extinction

Galactic (Milky Way) extinction is assumed to follow the $R_V = 3.1$ law typical of high Galactic latitude lines of sight. The reddening $E(B-V)$ is an input to the ETC. The extinction curve is obtained from Weingartner & Draine (2001) and Draine (2003a,b).[12]

For work in the visible and NIR and at high Galactic latitude, the *uncertainty* associated with the Galactic extinction correction is probably negligible.

### D. Line sensitivity and effective solid angle

This section describes the computations for emission line sensitivity in the spectrometer. Note however that the SNR computations for objects in the imager follow an analogous procedure, with the exceptions that (i) the 1st order throughput $f_1$ is replaced with the 0th order throughput $f_0$; (ii) one uses a polychromatic rather than monochromatic PSF; and (iii) there is no "source continuum" contribution (since for imaging the source continuum *is* the signal).

The limiting sensitivity to a line emitter depends on reaching a particular SNR. The "signal" in this case is the number of photons received from the galaxy. This is given by

$$\text{Signal} = \frac{N_{\text{exp}} \lambda A f_1 F t_1}{hc},$$

where $N_{\text{exp}}$ is the number of exposures; $\lambda$ is the observed wavelength of the line; $A$ is the geometric collecting area of the telescope; $f_1$ is the 1st order throughput, i.e. the fraction of the photons entering the geometric area that produce collected electrons in the 1st order spectrum; $F$ is the line flux from the galaxy; $t_1$ is the duration of each exposure; and $h$ and $c$ are fundamental constants. (In the BAO wavelength range, we do not expect events with multiple electrons per photon.)

The effective noise for a galaxy in the absence of source continuum is computed using:

---

[12] ftp://ftp.astro.princeton.edu/draine/dust/mix/kext_albedo_WD_MW_3.1_60_D03.all

$$\text{Noise}^2 = N_{\text{exp}} \frac{\Omega_{\text{eff}} \sigma_{\text{eff}}^2}{P^2},$$

where $\sigma_{\text{eff}}^2$ is the effective noise variance per pixel per exposure; $\Omega_{\text{eff}}$ is the effective solid angle of the source; and $P$ is the pixel scale. The effective noise has several contributions, as described in §7.F.

The effective solid angle for computing S/N ratios for the background-limited case (relevant here) depends on the normalized probability distribution $\Pi(\mathbf{x})$ for the apparent 2D angular positions $\mathbf{x}$ [rad] of line photons arriving from a galaxy. It is given by

$$\Omega_{\text{eff}} = \frac{1}{\int [\Pi(\mathbf{x})]^2 d^2\mathbf{x}} = \frac{1}{\int |\tilde{\Pi}(\mathbf{u})|^2 d^2\mathbf{u}},$$

where in the second equality we have considered the Fourier transform $\tilde{\Pi}(\mathbf{u})$, which depends on the spatial frequency vector $\mathbf{u}$. Note that we define the Fourier transform using spatial frequencies,

$$\Pi(\mathbf{x}) = \int \tilde{\Pi}(\mathbf{u}) e^{2\pi i \mathbf{u}\cdot\mathbf{x}} d^2\mathbf{u} \quad \Leftrightarrow \quad \tilde{\Pi}(\mathbf{u}) = \int \Pi(\mathbf{x}) e^{-2\pi i \mathbf{u}\cdot\mathbf{x}} d^2\mathbf{x}.$$

$\tilde{\Pi}(\mathbf{u})$ is given by the product of the Fourier transform of the normalized galaxy profile $\tilde{f}(\mathbf{u})$ and the Fourier transform of the PSF $\tilde{G}(\mathbf{u})$. "Normalization" here means that $\tilde{f}(\mathbf{0}) = \tilde{G}(\mathbf{0}) = \tilde{\Pi}(\mathbf{0}) = 1$. Also, we can see that for a top-hat distribution where the probability density is equal to $\Omega^{-1}$ (within some region of angle $\Omega$) and 0 (elsewhere), the effective area is $\Omega_{\text{eff}}=\Omega$, hence its name. Smaller solid angle objects (i.e. smaller galaxies or smaller PSFs) will have lower noise and hence we can expect to reach lower flux sensitivities in these cases.

In BAO mode, we restrict the integral over $\mathbf{u}$ to the first Brillouin zone ($|u_x|$, $|u_y|$ < 0.5/$P$), which provides the correct expected detection SNR for a large number of undersampled images naively sinc-interpolated onto a common grid and coadded. For practical WFIRST cases this restriction has little effect on the galaxy yields.

The expected SNR ("SNRe") is obtained as the ratio of the signal (calculated using the *true* line flux) to the noise. For the BAO, the default option is to impose cuts on the observed SNR ("SNRo"), which is the measured signal divided by the noise. The two are not identical (SNRe is a property of the galaxy and the instrument, whereas SNRo is a random variable); see §7.E.

The <u>statistical</u> uncertainty (<u>not</u> including aliasing effects) in the centroid in the $x_1$-direction is

$$\sigma_x^{-2} = I_C \frac{[\text{source counts}]}{[\text{background counts per unit solid angle}]},$$

where the centroiding integral is $I_C = \int \left|2\pi u_1 \tilde{\Pi}(\mathbf{u})\right|^2 d^2\mathbf{u}$. This is used in the routine `get_gal_centroid_integral`, but is not used for any code outputs at present.

### E. Expected vs. observed SNR

The BAO program often works at low SNR (e.g. a reasonable detection significance is 7σ), in which case we distinguish SNRe versus SNRo. When we analyze the actual data, the actual cut is on the measured SNRo. The computation of §7.D, however, yields SNRe. We must therefore describe the conversion between the two. In terms of galaxy yields, the difference between SNRe and SNRo is related to the phenomenon of Eddington bias – some galaxies scatter above the cut due to noise, and some scatter below, but generally the corrections do not cancel out and depend on the slope of the HαLF. As in the standard derivation of Eddington bias, we keep terms only through second order in the noise (Eddington 1913). The derivation is straightforward but given here to ensure that the underlying assumptions are clear.

The leading-order relation between SNRe (for optimal extraction) and SNRo can be obtained from elementary statistics. We consider an image of size $n$ pixels to form a data vector $|d\rangle \in \mathbf{R}^n$, and define an inner product of two vectors $|d\rangle$ and $|e\rangle$ based on the inverse-covariance matrix: $\langle d|e\rangle = \sum_{ij} [\mathbf{C}^{-1}]_{ij} d_i e_j$ (so that we have a real Hilbert space). We consider both the data and a template vector $|t(\lambda)\rangle$ corresponding to a line of unit flux (1 W m$^{-2}$) at observed wavelength λ. Then if the true line flux is $F_0$ and the true wavelength is $\lambda_0$, the expected SNR is

$$\mathrm{SNRe} = F_0 \sqrt{\langle t(\lambda_0)|t(\lambda_0)\rangle}.$$

For comparison, the observed SNR is

$$\mathrm{SNRo} = \max_\lambda \frac{\langle d|t(\lambda)\rangle}{\sqrt{\langle t(\lambda)|t(\lambda)\rangle}},$$

where the maximization is taken over possible observed wavelengths. The relation between these can be understood if we treat the data vector as the expected vector plus an error $|\varepsilon\rangle$ drawn from the covariance matrix, i.e. so that $\langle|\varepsilon\rangle\langle\varepsilon|\rangle = \hat{I}$ is the identity operator:

$$|d\rangle = F_0 |t(\lambda_0)\rangle + |\varepsilon\rangle.$$

We further note that $A^2 = \langle t(\lambda)|t(\lambda)\rangle$ depends very slowly on wavelength (the image of the spectral line is simply translated, so the SNRe of a line of given flux stays the same except for aliasing effects) and so treat it as constant. Then we find

$$\mathrm{SNRo} = \frac{1}{A} \max_\lambda \left[ F_0 \langle t(\lambda_0)|t(\lambda)\rangle + \langle \varepsilon|t(\lambda)\rangle \right].$$

As long as the maximum is near $\lambda_0$ (i.e. for correctly identified lines), we may Taylor-expand the terms in the maximum. To second order in $\varepsilon$ and $\Delta\lambda = \lambda - \lambda_0$, we find

$$\text{SNRo} = \frac{1}{A}\max_{\Delta\lambda}\begin{bmatrix} F_0\langle t(\lambda_0)|t(\lambda_0)\rangle + \langle\varepsilon|t(\lambda_0)\rangle + F_0\langle t(\lambda_0)|t'(\lambda_0)\rangle\Delta\lambda + \langle\varepsilon|t'(\lambda_0)\rangle\Delta\lambda \\ +\tfrac{1}{2}F_0\langle t(\lambda_0)|t''(\lambda_0)\rangle\Delta\lambda^2 \end{bmatrix}.$$

Here the primes denote the derivative of the template with respect to $\lambda$. In the brackets, the 3rd term vanishes and $\langle t(\lambda_0)|t''(\lambda_0)\rangle = -\langle t'(\lambda_0)|t'(\lambda_0)\rangle$.[13] Using the quadratic formula to find the maximum gives

$$\text{SNRo} = \frac{1}{A}\left[F_0\langle t(\lambda_0)|t(\lambda_0)\rangle + \langle\varepsilon|t(\lambda_0)\rangle + \frac{\langle\varepsilon|t'(\lambda_0)\rangle^2}{2F_0\langle t'(\lambda_0)|t'(\lambda_0)\rangle}\right].$$

We can then see that the mean[14] and variance of SNRo given SNRe are (again to 2nd order in $\varepsilon$):

$$\langle\text{SNRo}\rangle = \text{SNRe} + \frac{1}{2\text{SNRe}} \quad \text{and} \quad \text{Var}(\text{SNRo}) = 1.$$

In the high-SNR limit where $\varepsilon$ is a small perturbation, we can treat SNRo as a random variable with this mean and variance.

Now given a particular class of galaxies (e.g. H$\alpha$ emitters at a given redshift) there is some probability distribution per Mpc$^3$ $\phi(\text{SNRe})$, and correspondingly a cumulative distribution $\Phi(\text{SNRe}) = \int_{\text{SNRe}}^{\infty}\phi(x)dx$. We need to know the number density of galaxies with SNRo greater than some cut $Z$ (typically 7). To do this, we need to do a double integral over both SNRe and SNRo:

$$\begin{aligned} n &= \int_Z^{\infty}\int_0^{\infty}\phi(\text{SNRe})P(\text{SNRo}|\text{SNRe})\,\text{dSNRe}\,\text{dSNRo} \\ &= \int_{\mathbf{R}^2}\phi(\text{SNRe})P(\text{SNRo}|\text{SNRe})\Theta(\text{SNRo}-Z)\,\text{dSNRe}\,\text{dSNRo} \\ &\approx \int_{\mathbf{R}}\phi(\text{SNRe})\left[\Theta(\text{SNRe}+\tfrac{1}{2}\text{SNRe}^{-1}-Z) + \tfrac{1}{2}\Theta''(\text{SNRe}+\tfrac{1}{2}\text{SNRe}^{-1}-Z)\right]\text{dSNRe} \\ &= \Phi(x) + \tfrac{1}{2}\int_{\mathbf{R}}\phi(\text{SNRe})\delta'(\text{SNRe}+\tfrac{1}{2}\text{SNRe}^{-1}-Z)\,\text{dSNRe}, \end{aligned}$$

where $x$ is the value of SNRe satisfying $\text{SNRe}+\tfrac{1}{2}\text{SNRe}^{-1}-Z=0$ (to second order in the noise: $x = Z - \tfrac{1}{2}Z^{-1}$), $\Theta$ is the Heaviside step function, and $\delta$ is the Dirac impulse function.[15]

---

[13] This can be seen by taking the first and second derivatives of the $A^2$ equation.
[14] Note that the mean of SNRo is *greater than* SNRe even for unbiased data, by 1/(2SNRe) for each parameter (aside from the amplitude) in the fit – in this case we included only the 1 parameter $\lambda$. This bias is little appreciated by many astronomers but is routinely corrected in some large surveys – see e.g. Condon *et al.* (1998), §5.2.5.

In the third line we have used the moments of the SNRo distribution and performed a formal Taylor expansion of Θ through 2nd order. The remaining integral can be reduced via the chain rule and then integration by parts to

$$\begin{aligned} n &= \Phi(x) + \tfrac{1}{2} \int_{\mathbf{R}} \phi(\text{SNRe}) \delta'(\text{SNRe} + \tfrac{1}{2}\text{SNRe}^{-1} - Z) \, d\text{SNRe} \\ &= \Phi(x) + \tfrac{1}{2} \int_{\mathbf{R}} \frac{\phi(\text{SNRe})}{1 + \tfrac{1}{2}\text{SNRe}^{-2}} \frac{d}{d\text{SNRe}} \delta(\text{SNRe} + \tfrac{1}{2}\text{SNRe}^{-1} - Z) \, d\text{SNRe} \\ &= \Phi(x) - \tfrac{1}{2} \left. \frac{d}{d\text{SNRe}} \frac{\phi(\text{SNRe})}{1 + \tfrac{1}{2}\text{SNRe}^{-2}} \right|_{\text{SNRe}=x} \\ &= \Phi(x) - \tfrac{1}{2} \frac{\phi'(x)}{1 + \tfrac{1}{2}x^{-2}} - \tfrac{1}{2} \frac{x^{-3}\phi(x)}{\left(1 + \tfrac{1}{2}x^{-2}\right)^2}. \end{aligned}$$

Now if we keep only the fractional corrections to $n$ that are of order $Z^{-2}$, we may write this as $n \approx \Phi(x) - \tfrac{1}{2}\phi'(x)$. To this order, we can then write the finite difference equation

$$n \approx \Phi(x) + \tfrac{1}{2}\Phi''(x) \approx \tfrac{1}{2}\left[\Phi(x+1) + \Phi(x-1)\right] \approx \tfrac{1}{2}\left[\Phi(Z+1-\tfrac{1}{2}Z^{-1}) + \Phi(Z-1-\tfrac{1}{2}Z^{-1})\right].$$

Thus we conclude that to leading order in the noise terms (i.e. order $Z^{-2}$), the number density of objects with SNRo>Z can be computed as the arithmetic average of the number with SNRe greater than $Z+1-\tfrac{1}{2}Z^{-1}$ and greater than $Z-1-\tfrac{1}{2}Z^{-1}$. This provides a simple approach to the SNRo issue for general luminosity functions, without the need for numerical or analytical derivatives of the HαLF.

The above model will give nonsensical results if $Z < 1.37$ because the asymptotic series in $1/Z$ is ill-behaved; in practice using it for $Z < 4$ (or using thresholds this low!) is unadvisable and a warning will be given.

### F. <u>Noise model</u>

The noise model currently contains sky noise (zodiacal light), dark current, and read noise. The computation has two parts: the determination of the expected counts in the detector, and the determination of the effective noise $\sigma_{\text{eff}}^2$ including both Poisson and read contributions.

The expected number of counts per pixel $Q$ has several sources. First is the sky background:

$$Q_s = t_1 A P^2 \int_{\lambda_{\min}}^{\lambda_{\max}} \frac{\lambda I_{\lambda,\text{sky}}}{hc} f_{\text{all}}(\lambda) d\lambda,$$

Here $I_{\lambda,\text{sky}}$ is the sky brightness in units of power per unit area per unit solid angle per unit wavelength; and $f_{\text{all}}(\lambda)$ is the all-order throughput, i.e. the fraction of the photons entering the geometric area that produce collected electrons in *any* order [note: generally $f_{all}(\lambda) \geq f_1(\lambda)$]. There is a simpler expression for the dark counts:

$$Q_d = I_d t_1,$$

where the dark current $I_d$ has units of $e^-$/s/pix.

---

[15] The primes in this equation denote derivatives with respect to the argument, not with respect to SNRe.

There are several options for incorporating the read noise. For the default (NIR detectors) the code currently accepts a read noise floor $N_{read}$ from the user. The default behavior is to assume an unweighted SUTR fit:[16]

$$\sigma_{eff}^2 = \frac{6(n^2+1)}{5n(n^2-1)} \frac{t_1}{t_f} Q + \frac{12}{n(n^2-1)} \left(\frac{t_1}{t_f}\right)^2 \sigma_{read}^2 + N_{read}^2,$$

where the number of frames is $n$, the read noise variance per frame is $\sigma_{read}^2$, and the total charge is $Q$. The number of frames is computed as $n$ = `floor`($t_{exp}/t_f$), where `floor` rounds down to the next integer.[17] The time per frame $t_f$ (in seconds) and $\sigma_{read}^2$ are currently set by the global variables `T_FRAME` and `VAR_READ_FRAME`, which are set at the beginning of a run when the user provides a detector type.

The default behavior can be changed by several options:

- The `-DLOGICAL_READ_FLOOR` option takes only the larger of the two read noise related terms in $\sigma_{eff}^2$.
- If a CCD detector is specified, the ETC simply adds the Poisson variance to the read noise floor: $\sigma_{eff}^2 = Q + N_{read}^2$. This is not appropriate for WFIRST, but is provided to model the Euclid visible channel (or any other CCD-based mission that we may wish to analyze).

In the BAO mode, there is an additional contribution to the noise associated with Poisson fluctuations in the source continuum. This additional contribution is a function of position in the image, being greatest on the centerline of the source spectral trace and going to zero far away from it. This additional contribution is

$$\sigma_{eff,src}^2(x_1, x_2) = f_d P^2 \frac{A f_1 F_\nu t_1}{h D_\theta} \int \Pi(x_1', x_2) dx_1',$$

where $F_\nu$ is the source continuum flux (in W m$^{-2}$ Hz$^{-1}$), $D_\theta$ is the spectral dispersion (in arcsec per $\Delta \ln \lambda$), $(x_1, x_2)$ are spatial coordinates in the dispersion and cross-dispersion directions, and $f_d$ is the Poisson noise enhancement factor (equal to 1 for a CCD and 6/5 for a long exposure on a NIR detector; the ETC uses 6/5 for all NIR cases since this is a relatively minor term). In the BAO mode, we care about the average value of the noise variance weighted over the image of the spectral line, i.e.:

$$\sigma^2_{\text{eff,src-avg}} = \int \sigma^2_{\text{eff,src}}(\mathbf{x}) \Pi(\mathbf{x}) \, d^2\mathbf{x} = f_d P^2 \frac{A f_1 F_v t_1}{h D_\theta} J,$$

where the overlap integral $J$ is given by

$$J = \int \Pi(x_1, x_2) \Pi(x'_1, x_2) \, dx'_1 \, dx_1 \, dx_2 = \int |\tilde{\Pi}(0, u_2)|^2 \, du_2.$$

---

[16] This is equivalent to Eq. (1) of Rauscher *et al.* (2007), without grouping ($m$ = 1 in their notation). The normalization has been re-scaled to estimate the uncertainty in the number of counts in time $t_1$ instead of $(n-1)t_f$.

[17] An error message is returned if $n < 2$ since in this case the SUTR fit is impossible. This is not of practical importance for any plausible WFIRST observing strategy.

This has units of arcsec$^{-1}$. Its conceptual interpretation is that it converts the source continuum brightness (units: counts per linear arcsec along the trace) to an effective background (counts per arcsec$^2$) that has the same impact on the SNR. The second integral (in terms of the Fourier transform) is actually implemented in the code. For a Gaussian, we would have $J = 0.33/\text{EE50}$, but the code does the full integral (whose value in realistic cases tends to be of order ~10% smaller).

Note that for a line of *fixed equivalent width* the noise variance depends linearly on the line flux. Therefore the limiting sensitivity requires the solution of an equation with the line flux on both sides. This equation is solved in `get_limflux`; it is a quadratic equation and presents no special difficulty.

The calculation requires the product of rest-frame equivalent width EW$_{\text{rest}}$ and dispersion:

$$\frac{F_v}{D_\theta} = \frac{\lambda}{c[\text{EW}_{\text{rest}} D_\theta / \lambda_{H\alpha}]} F.$$

We have set the product in brackets equal to 3.656 arcsec, which is appropriate for Hα rest frame equivalent width of 100 Å and dispersion $D_\theta$ = 240 arcsec. This is set in the preamble as:

```
/* Reference equivalent width in arcsec -- 100A EW rest frame @
   D_theta = 240 arcsec. */
#define EWD_REF 3.656
```

If the −DNO_OBJ_CONTINUUM flag is set, then the object in brackets is set to "∞" (technically $10^{12}$ arcsec), which turns object continuum off.

### G. Thermal background model

Thermal emission from the telescope may be a significant noise contribution for WFIRST. It can come from several sources, including:

1. The "telescope" (primary and secondary mirrors and associated structures).

2. Additional optics upstream of the filter, such as the tertiary mirror and fold flats; and
3. Other components, such as the detector enclosure, cold mask, etc., as well as the filter and any optical components downstream of the filter.

Thermal emission from components upstream of the filter (items 1 and 2 of this list) is suppressed by the filter transmission curve. The ETC computes thermal emission from each source in $e^-$/pix/s, and then adds them up.

The telescope is modeled as a single thermal zone with temperature $T_{tel}$. The starting point for the computation is the flux of photons $F_{bb}$ (units: photons µm$^{-2}$ s$^{-1}$ sr$^{-1}$) from a blackbody observed through the throughput + quantum efficiency curve of the optics + detector Thr($\lambda$).[18] This is given by the usual blackbody formula

$$F_{bb}(T) = \int_{\lambda_{min}}^{\lambda_{max}} \frac{2c}{\lambda^4 (e^{hc/kT\lambda} - 1)} \text{Thr}(\lambda) \, d\lambda.$$

If the telescope looked into a blackbody at temperature $T_{tel}$, and one neglected emission from all optics downstream of the secondary, then there would be a consequent flux of detected photons per pixel per second per steradian given by

$$\frac{dI}{d\Omega} = \frac{T_{filter} F_{bb}(T_{tel}) \Delta^2}{(1-\varepsilon)^2},$$

where $T_{filter}$ is the filter transmission in band, $\Delta$ is the physical pixel scale (in µm; this is equal to $P$ times the effective focal length), and $\varepsilon$ is the mirror emissivity. The factor of $(1-\varepsilon)^2$ is in practice a minor contribution but is in principle necessary since in the aforementioned situation there is blackbody radiation emerging from the secondary mirror, but the net throughput curve through which it is seen is larger than Thr($\lambda$) by this factor.

Finally, one desires the effective emissivity-weighted solid angle for which this emission applies. The effective emissivity is taken as 1 for rays that emerge from "black" components such as baffles, and is $2\varepsilon - \varepsilon^2$ for rays that emerge from the sky and bounce off both mirrors. For an ideal circular aperture the solid angle would be $\pi/(4f^2)$, where $f$ is the effective f/ratio. We thus model the emissivity-weighted solid angle for the telescope emission as

$$\Omega_t = \frac{\pi}{4f^2}(2\varepsilon - \varepsilon^2)(1 - \upsilon^2)$$

(where $\upsilon$ is the linear central obscuration) for cases with a pupil mask, and

$$\Omega_t = \frac{\pi}{4f^2}\left[(2\varepsilon - \varepsilon^2)(1-\upsilon^2) + \upsilon^2 + \rho^2 - 1\right]$$

---

[18] The function `get_BB_flux_filter`, which computes $F$, assumes the filter to have perfect emission in band – we account for the finite in band transmission in the functions that call `get_BB_flux_filter`.

(where $\rho \geq 1$ is the linear ratio of the beam size that passes through the limiting mask at the exit pupil to the actual size of the primary mirror) for cases without a pupil mask. Here the term $\upsilon^2 + \rho^2 - 1$ accounts for emission from the black components: the $\upsilon^2$ is from rays that are blocked by the central obscuration and $\rho^2 - 1$ is due to over-illumination of the primary.[19]

The telescope contribution to the thermal background (in $e^-$/pix/s) is then

$$I_t = \frac{dI}{d\Omega}\Omega_t.$$

The next background that we consider is that from additional optics upstream of the filter (e.g. the fold flats and tertiary). These are taken to have a net emissivity $\varepsilon_{aft}$ – this is the total for all of the optical components between the secondary and filter, e.g. $\varepsilon_{aft} = 1-(1-0.02)^4 = 0.0776$ for a sequence of 4 mirrors each of emissivity 0.02. It is assumed that this emission fills the full solid angle admitted by any mask at the exit pupil. The aft optics contribution is then:

$$I_{aft} = \frac{\pi\rho^2}{4f^2}\frac{\varepsilon_{aft}T_{filter}}{(1-\varepsilon_{aft})(1-\varepsilon)^2}F_{bb}(T_{aft})\Delta^2$$

for the case without a pupil mask; with the mask the factor of $\rho^2$ is removed.

Finally, we consider a contribution associated with all structures downstream of the filter. This contribution is usually small and we simply assume a specification that it will not exceed some value $I_{down}$.

A margin factor `THERMAL_MARGIN_FACTOR` may in principle be assigned. This is currently set to 1.00 (margin is <u>not</u> applied within the thermal backgrounds analysis, and is assumed to be applied elsewhere).[20] The overall thermal current in $e^-$/pix/s is then

$$I_{therm} = \text{THERMAL\_MARGIN\_FACTOR}(I_t + I_{aft} + I_{down}).$$

The default values of the parameters are as follows. Note that these can be changed with a `!THERMAL` line in the telescope configuration file.

- Temperatures for telescope and aft optics: $T_{tel} = T_{aft} = 230$ K.
- Cold pupil mask not included, $\rho = 1.05$.
- Emissivity per surface $\varepsilon = 0.02$, aft optics have 4 reflections $\varepsilon_{aft} = 0.0776$.
- Downstream component thermal background $I_{down} = 0.005$ $e^-$/pix/s (imager) or 0.03 $e^-$/pix/s (spectrometer).

---

[19] Intermediate cases can also be implemented. For example, an outer pupil mask only on an obstructed telescope would be modeled by using the "no mask" formula with $\rho=1$, $\upsilon>0$.
[20] This is controlled by the macro `#define THERMAL_MARGIN_FACTOR 1.00`.

Note that dark current is <u>not</u> included in the thermal background model (even if thermal in origin) – this is included separately.

### H. **Galaxy profiles**

The galaxy profiles will be taken to be exponential, as this is characteristic of late-type galaxies that host emission lines. A normalized exponential profile with half-light radius $r_{eff}$ [rad] is given by:

$$f(\mathbf{x}) = \frac{1}{2\pi r_s^2} \exp\frac{-|\mathbf{x}|}{r_s}.$$

The scale radius can be found numerically to be $r_s = r_{eff} / 1.67834$. The Fourier transform is

$$\tilde{f}(\mathbf{u}) = \left(1 + 4\pi^2 r_s^2 |\mathbf{u}|^2\right)^{-3/2}.$$

### I. **Point spread functions**

The ETC computes the monochromatic MTF $\tilde{G}(\mathbf{u})$ directly from the autocorrelation of the aperture. This autocorrelation is computed analytically for an annulus using elementary geometry. The MTF is further modified by jitter and charge diffusion. These may in general have complicated forms, but for the purposes of the ETC they are modeled by a Gaussian of width σ per axis: $\tilde{G}(\mathbf{u}) := \tilde{G}(\mathbf{u})\exp(-2\pi^2\sigma^2|\mathbf{u}|^2)$. The jitter is provided by the user, and the charge diffusion is set by the global `PIX_CD` parameter (selected for each detector type). Finally, the pixel tophat is included, $\tilde{G}(\mathbf{u}) := \tilde{G}(\mathbf{u})\mathrm{sinc}(\pi P u_x)\mathrm{sinc}(\pi P u_y)$.

Aberrations cannot be fully specified in the current version of the ETC. They are included by multiplying the MTF by a fitting function:

$$\tilde{G}_{\text{true}}(\mathbf{u}) = \tilde{G}_{\text{no aberrations}}(\mathbf{u})\exp\left\{-72\frac{\text{RMSWFE}^2}{\lambda^2}\left[1 - \tfrac{1}{2}e^{-8(\lambda\mathbf{u}/D)^2} - \tfrac{1}{2}e^{-32(\lambda\mathbf{u}/D)^2}\right]\right\}.$$

This has been tested against the "correct" MTFs (computed by FFT of an aperture with the appropriate phase errors) for several possible aberrations including defocus, astigmatism, coma, and trefoil, and various combinations thereof, and for RMSWFE = 0.05λ or 0.075λ. Since some of these aberrations are not circularly symmetric, the relevant point of comparison (for WL shapes) is the effect on the isotropically averaged power spectrum of the noise in the sky maps, which depends on the quantity:

$$f(u) = \frac{1}{2\pi}\int_0^{2\pi}\frac{d\varphi}{|\tilde{G}(u\cos\varphi, u\sin\varphi)|^2}.$$

For this reason, we compare to the "noise-effective MTF" $\tilde{G}_{\rm eff}(u) = [f(u)]^{-1/2}$; in each of the cases considered, the fitting formula used was either conservative or within <0.5%. An example of such a comparison is shown in Figure 1.

Emission line detection depends on the angular rms of the MTF,

$$\tilde{G}_{\rm rms}(u) = \sqrt{\frac{1}{2\pi}\int_0^{2\pi}\left|\tilde{G}(u\cos\varphi,u\sin\varphi)\right|^2 d\varphi},$$

rather than the noise-effective MTF. This has been tested against the same combinations of aberrations in Figure 1 up to RMSWFE = 0.125λ.

At RMSWFE = 0.175—0.225λ, the above fitting function is no longer conservative if the dominant aberration is defocus. Thus in this regime, we switch to the fitting formula

$$\tilde{G}_{\rm true}(\mathbf{u}) = \tilde{G}_{\rm no\,aberrations}(\mathbf{u})\frac{\exp\left(-2160{\rm RMSWFE}^2\mathbf{u}^2/D^2\right)+\exp\left(-720{\rm RMSWFE}^2\mathbf{u}^2/D^2\right)}{2},$$

which is conservative except for a very small region in the case of pure defocus where $\tilde{G}_{\rm rms}(u)$ passes through a null (see Figure 2; the overall effect is in the direction of being conservative in terms of evaluating the **u** integrals even in this case). Between 0.125λ and 0.175λ we linearly interpolate between the two formulae.

An alternative model, which can be specified by the -DWFE_HOPKINS_EXP flag, uses an exponential model for the Hopkins ratio (Olson 2008) instead:

$$\tilde{G}_{\rm true}(\mathbf{u}) = \tilde{G}_{\rm no\,aberrations}(\mathbf{u})\exp\left[-4\pi^2\frac{{\rm RMSWFE}^2}{\lambda^2}\left(1-e^{-18(\lambda\mathbf{u}/D)^2}\right)\right].$$

This model has a wavefront error with a Gaussian correlation function of scale length $D/3$, as assumed during the JDEM-ISWG studies (Levi *et al*. 2011).

The PSF routines return an error message if they are used outside the tested range of RMSWFE.

The WL calculations require a polychromatic MTF. The ETC generates the polychromatic MTF from 12 monochromatic MTFs spaced logarithmically in wavelength across the bandpass, and then does an arithmetic average. This generates the correct polychromatic MTF for a constant AB magnitude source with λ-independent throughput. The small variations with source color must be considered in data analysis, but will not have a significant effect on the exposure time calculation.

The BAO calculations include a contribution added to the jitter to account for the smearing of the galaxy by the finite spectral resolution. This contribution is

$$\sigma = \frac{D_\theta \sigma_v}{\sqrt{2}\,c},$$

and is added in quadrature to the jitter before computing the PSF. Here $D_\theta$ is the spectral dispersion, $\sigma_v$ is the 1D rms velocity dispersion of the galaxy, and $c$ is the speed of light. The factor of $\sqrt{2}$ accounts for the smearing in only one of the two dimensions. The value of this contribution to σ is controlled by the macro:

```
#define SIGMA_PROF 0.040
```

This value is appropriate for $D_\theta$ = 240 arcsec and $\sigma_v$ = 70 km/s, which are typical values for WFIRST and for an emission line galaxy, respectively. The behavior of this correction is not a perfect representation since real line profiles are non-Gaussian, they vary across the image of the galaxy, and the induced pattern is not circularly symmetric. Nevertheless, since this term typically affects galaxy yields at only the ~1% level, we believe the simplified treatment is adequate for the purpose of an ETC.

We have found that of the aberrations considered, at a fixed RMSWFE, the most damaging for the WL is the 50:50 mix of defocus and astigmatism (this also yields the largest PSF ellipticity) and the most damaging for the BAO is pure coma.

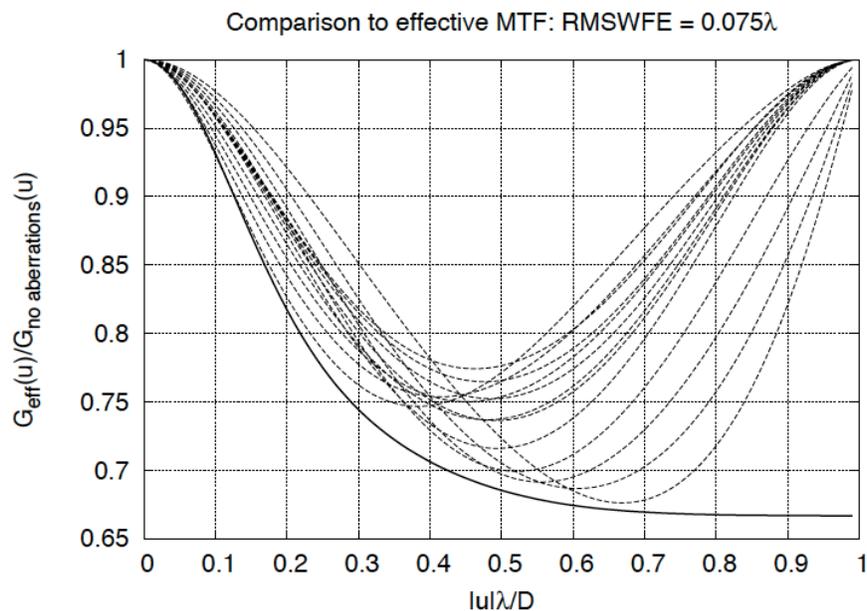

**Figure 1: The effective MTF for an unobstructed aperture with aberrations, relative to the unaberrated case, at RMSWFE = 0.075λ. The solid curve is the fitting formula used in the ETC. The dashed curves are low-order aberrations. There are 13 cases shown, with different aberrations: defocus, astigmatism, coma, and trefoil; and cases with 25:75, 50:50, and 75:25 mixtures of defocus with the other three aberrations.**

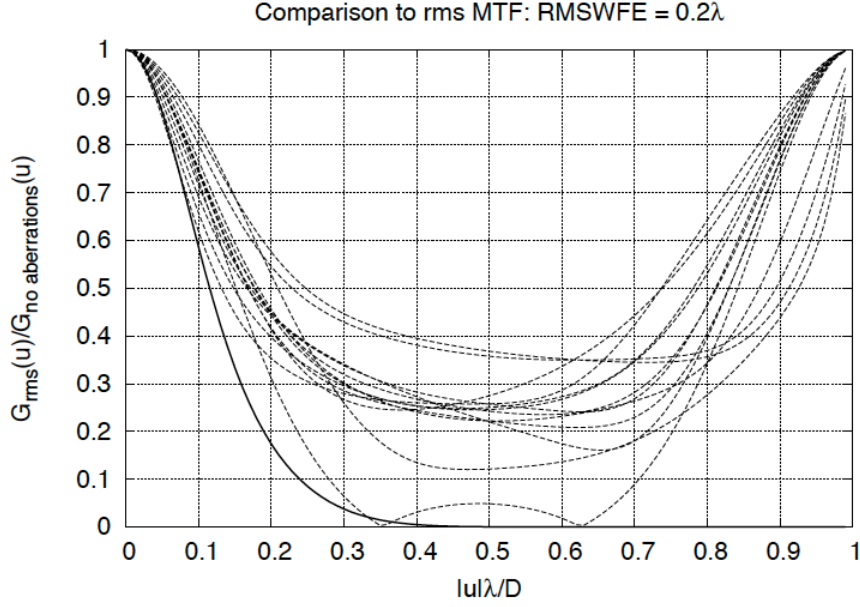

**Figure 2:** The rms MTF for an unobstructed aperture with aberrations, relative to the unaberrated case at RMSWFE = 0.2λ. The solid curve is the fitting formula used in the ETC. The dashed curves are low-order aberrations. There are 13 cases shown, with different aberrations: defocus, astigmatism, coma, and trefoil; and cases with 25:75, 50:50, and 75:25 mixtures of defocus with the other three aberrations.

## J. Ellipticity uncertainty

The uncertainty in the ellipticity is estimated by standard techniques (Bernstein & Jarvis 2002, §3). The result is that rather than the effective area $\Omega_{\rm eff}$, the uncertainty in the ellipticity depends on the shape-effective area $\Omega_{\rm eff,s}$:

$$\Omega_{\rm eff,s} = \frac{1}{\int \left| \tilde{G}(\mathbf{u}) \partial_{\ln u} \tilde{f}(\mathbf{u}) \right|^2 d^2\mathbf{u}}.$$

Then the ellipticity measurement uncertainty $\sigma_e$ is 2 divided by the signal-to-noise ratio that one would obtain for a galaxy of the usual flux occupying solid angle $\Omega_{\rm eff,s}$. One may define the penalty factor:

$$f_{\rm pen} = \frac{\Omega_{\rm eff,s}}{\Omega_{\rm eff}};$$

the formula for the ellipticity error then reads $\sigma_e = 2f_{\rm pen}^{1/2}/{\rm SNR}$. The penalty factor is 1 for a well-resolved Gaussian, and $R^{-2}$ (where $R$ is the resolution factor) for general Gaussian PSFs and galaxies. It always approaches ∞ for point sources. Note that for non-Gaussian profiles, the penalty factor can be oddly behaved even if the galaxy is well-resolved: peaked profiles such as the exponential have $f_{\text{pen}} > 1$ even in this limit. Conversely, objects with sharp edges can have $f_{\text{pen}} < 1$.

# 8. Data structures and functions

This section provides a complete listing of the internal data structures and functions in the code (including low-level utilities). It may be of use to users who wish to write their own driver routines.

### A. Data structures

There are three data structures currently in use: one to describe a PSF; one to describe throughput curves; and one to describe telescope thermal parameters.

```
typedef struct {
  double pixscale;   /* pixel scale in arcsec */
  double sigma;      /* Gaussian smearing, in arcsec */
  double lD;         /* central lambda/D * 206265 */
  double centobs;    /* central obscuration as a fraction of primary
          diameter */
  int is_broadband;  /* 0=monochromatic, 1=broadband */
  double dll;        /* fractional width of band, i.e. ln(lambda) =
          central +/- 0.5*dll;
                      * only used if broadband */
  double rmswfe;     /* rms wave front error in waves */
}
PSF_DATA;
```

Note that here `lD` ("$\lambda/D$") is measured in arcsec – e.g. for a 2.0 m telescope at 1.5 µm, `lD` = 0.1547. The flag `is_broadband` indicates a monochromatic (0, for spectral lines) or broadband (1, for star or galaxy continuua) PSF. For broadband PSFs, `lD` is defined at the geometric center of the bandpass, i.e. one sets:

$$\mathtt{lD} = \frac{\sqrt{\lambda_{\min}\lambda_{\max}}}{D} \quad \text{and} \quad \mathtt{dll} = \ln\frac{\lambda_{\max}}{\lambda_{\min}}.$$

The rms wavefront error is measured in waves at the geometric central wavelength $\sqrt{\lambda_{\min}\lambda_{\max}}$.

```
#define N_THROUGHPUT_MAX 512
typedef struct {
  int N;                                /* Number of points in
          interpolation grid */
  double lambda[N_THROUGHPUT_MAX];      /* Wavelengths (in microns) of
          nodes */
  double throughput[N_THROUGHPUT_MAX];  /* Throughput values */
}
```

```
THROUGHPUT_DATA;
```

The throughput curve is defined by linear interpolation through `N` points specified by their wavelength and throughput.

```
typedef struct {
  double T_tel;                   /* Telescope temperature [K] -- PM/SM +
          assoc struct */
  double pmsm_emissivity;         /* Emissivity of primary & secondary
          mirror surfaces (epsilon) */
  double outer_ann_ratio;         /* Ratio of outer diameter of beam to
          that of primary (rho) */
  int ismask;                     /* Cold pupil mask? 1 = yes, 0 = no */
  double T_aft;                   /* Temperature of aft optics [K] */
  double aft_net_emissivity;      /* Net emissivity of aft optics (total
          of all surfaces in 'series') */
  double  post_filter_rate;       /* Count rate in e-/pix/s from
          structures downstream of the filter */
}
THERMAL_DATA;
```

This structure contains all information required to compute the thermal emission that is not otherwise specified in the telescope data – see §7.G.

### B. Interface functions

```
void wait_for_user(void);
```

If -DKEEP_WINDOW_OPEN is set, waits for the user to hit the <return> key.

### C. Special functions

```
double getJ0(double x);
double getJ1(double x);
double geterf(double x);
```

These are the standard special functions $J_0(x)$, $J_1(x)$, and erf $x$. The Bessel functions are accurate to several parts in $10^8$, and the error function to much better accuracy.[21]

### D. Routines to set up detector parameters

```
void ConfigDetector(int det_type);
```

---

[21] The Bessel functions are implemented using a combination of the integral form at $|x|<3$ and the formulae of Abramowitz & Stegun (1964), Eqs. 9.4.3 & 9.4.6. The error function uses either the Taylor expansion or an integral transformation described in the comments.

This routine assigns values to the global variables describing the detectors: `PIXSIZE_PHYS`, `PIX_CD`, and (for NIR detectors) `T_FRAME` and `VAR_READ_FRAME`. It takes as input a detector type: the current options are specified in §3.C.

### E. Routines to compute the throughput

```
double get_throughput(double lambda, THROUGHPUT_DATA *T);
```

Obtains the throughput from a `THROUGHPUT_DATA` structure at the specified wavelength $\lambda$ (in µm). Returns 1 if `T` is null.

### F. Routines to compute the modulation transfer function (MTF)

```
double get_MTF_mono(double u, double v, PSF_DATA *psf);
```

Obtains the monochromatic MTF including detector effects at spatial frequency ($u$,$v$) in cycles/arcsec and using the PSF configuration in structure `psf`.

```
double get_MTF(double u, double v, PSF_DATA *psf);
```

Obtains the full MTF including detector effects at spatial frequency ($u$,$v$) in cycles/arcsec and using the PSF configuration in structure `psf`. If `psf` is monochromatic, this function acts as a wrapper to `get_MTF_mono`. If `psf` is polychromatic, then a superposition of `N_LAMBDA_INTEG` monochromatic PSFs with weights logarithmically spaced in $\lambda$ is generated. (By default, `N_LAMBDA_INTEG` = 12.)

### G. Routines to compute galaxy image properties

```
double get_gal_fracE(double r, double reff, PSF_DATA *psf);
```

Returns the fraction of energy (between 0 and 1) encircled in radius $r$ for an exponential profile galaxy with effective radius $r_{\rm eff}$ convolved with `psf`. All radii are in arcsec.

```
double get_gal_size(double eFrac, double reff, PSF_DATA *psf);
```

Returns the radius in arcsec containing a fraction eFrac (between 0 and 1) of the light from an exponential profile galaxy with effective radius $r_{\rm eff}$ convolved with `psf`. This is the inverse function of `get_gal_fracE`.

```
double get_gal_Omeff(double reff, PSF_DATA *psf);
```

Returns the effective solid angle $\Omega_{\rm eff}$ of an exponential profile galaxy with effective radius $r_{\rm eff}$ convolved with `psf`. All radii are in arcsec, and the solid angle is in arcsec$^2$.

```
double get_J_couplingIntegral(double reff, PSF_DATA *psf);
```

Computes the overlap integral $J$ (units: arcsec$^{-1}$) for an exponential galaxy with effective radius $r_{\rm eff}$ convolved with `psf`. This is needed to assess source continuum noise in the galaxy emission line extraction.

```
double get_gal_centroid_integral(double reff, PSF_DATA *psf);
```

Returns the centroiding information integral $I_C$ for a galaxy in arcsec$^{-4}$.

```
double get_shape_penalty_factor(double reff, PSF_DATA *psf);
```

Computes the shape measurement penalty factor $f_{\rm pen}$, i.e. the additional factor by which the noise variance must be reduced in order to measure a galaxy's ellipticity as well as that of a well-resolved Gaussian object with the same SNR. The galaxy is taken to have an exponential profile with radius $r_{\rm eff}$ arcsec.

```
double get_1sigma_flux_1exp(double reff, PSF_DATA *psf, double
   var_1exp, double calib_1exp);
```

Returns the 1σ uncertainty in the line flux (in W m$^{-2}$) from a galaxy for a single exposure, as a function of $r_{\rm eff}$ (in arcsec; assumes exponential profile), the PSF, noise variance per exposure (in electrons$^2$ arcsec$^{-2}$), and the calibration (electrons per exposure per W m$^{-2}$). This function assumes no source continuum contribution, and as such is deprecated for BAO purposes (except for testing or in cases where the source continuum is completely negligible) – you should use `get_limflux` for practical computations.

```
double get_limflux(double reff, PSF_DATA *psf, double var_1exp, double
   calib_1exp, double EWD, double Nexp, double Z);
```

Returns the flux limit (in W m$^{-2}$) from a galaxy in $N_{\rm exp}$ exposures, as a function of $r_{\rm eff}$ (in arcsec; assumes exponential profile), the PSF, noise variance per exposure (in electrons$^2$ arcsec$^{-2}$), and the calibration (electrons per exposure per W m$^{-2}$). Assumes a significance cut at $Z\,\sigma$, and an equivalent width-to-dispersion ratio (in arcsec) of `EWD`.

```
double get_maglim_shape(double r_eff, PSF_DATA *psf, int N_exp, double
   var_1exp, double calib_1exp, double max_ellip_err);
double get_maglim_shape_noSNRcut(double r_eff, PSF_DATA *psf, int
   N_exp, double var_1exp, double calib_1exp, double max_ellip_err);
```

Returns the limiting AB magnitude for shape measurements, as a function of $r_{\rm eff}$ (in arcsec; assumes exponential profile), the PSF, the noise variance per exposure (in electrons$^2$ arcsec$^{-2}$), the calibration (in electrons per exposure for an AB magnitude 0 source), and the maximum ellipticity error $\sigma_{e,\rm max}$. The function `get_maglim_shape` imposes cuts on the SNR as well as on $\sigma_e$. The function `get_maglim_shape_noSNRcut` imposes only the cut on $\sigma_e$.

## H. Routines to compute cosmological information

```
double computeDistance(double z);
double computeHubble(double z);
```

The comoving distance $D(z)$ and Hubble rate $H(z)$ in Mpc and $c$/Mpc respectively. (Note that the background cosmological parameters are hard-coded.)

### I. Routines to compute properties of the BAO galaxy distribution

```
void get_galsizes(double z, double FHa, double *med_reff, double
   *dispersion, int model);
```

For H$\alpha$ emitters at redshift $z$ and flux $F_{H\alpha}$ (in W m$^{-2}$), and using the specified galaxy model, returns the median $r_{eff}$ (in arcsec) and the 1$\sigma$ dispersion in ln $r_{eff}$. The code currently assumes the distribution to be lognormal.

```
double get_HaLF(double z, double FHa, int model);
void print_HaLF_model(FILE *fp, int model);
```

The function `get_HaLF` obtains the H$\alpha$LF at redshift $z$ and flux $F_{H\alpha}$ (in W m$^{-2}$), and using the specified galaxy model. The result is in galaxies per comoving Mpc$^3$ per $\Delta$ ln $F_{H\alpha}$. The function `print_HaLF_model` outputs to the filestream the choice of H$\alpha$LF.

### J. Routines to compute densities of observable galaxies: BAO

```
double get_P_Detectable(double  z, double  FHa, PSF_DATA *psf, double
   var_1exp, double calib_1exp, int N_exp, double significance_cut, int
   model);
double get_n_Detectable(double  z, double  FHa, PSF_DATA *psf, double
   var_1exp, double calib_1exp, int N_exp, double significance_cut, int
   model);
```

The routine `get_P_Detectable` gets the probability that a galaxy at redshift $z$ and flux $F_{H\alpha}$ would be detectable at `significance_cut` $\sigma$. This probability is nontrivial (i.e. not just a flux cut) due to the distribution of $r_{eff}$ (obtained from `get_galsizes`). The routine requires ancillary data: the PSF, noise variance per exposure (in electrons$^2$ arcsec$^{-2}$), and the calibration (electrons per exposure per W m$^{-2}$), and number of exposures. The routine `get_n_Detectable` translates this into a number of detectable galaxies per comoving Mpc$^3$ per $\Delta$ ln $F_{H\alpha}$. If -DUSE_SNRE is set, then `get_n_Detectable` is simply `get_P_Detectable` times the H$\alpha$LF. If -DUSE_SNRE is not set (default), then `get_n_Detectable` includes a correction for the conversion from SNRe to SNRo. (That is, `get_P_Detectable` always cuts on the expected SNR, but `get_n_Detectable` cuts on either the expected or observed SNR as requested by the user.)

```
double get_n_galaxies(double z, PSF_DATA *psf, double var_1exp, double
   calib_1exp, int N_exp, double significance_cut, int model, double
   *stats);
```

This routine returns the commoving density of detectable galaxies at redshift $z$ and flux $F_{H\alpha}$ would be detectable at `significance_cut` σ. The routine requires ancillary data: the PSF, noise variance per exposure (in electrons$^2$ arcsec$^{-2}$), and the calibration (electrons per exposure per W m$^{-2}$), and number of exposures. Ancillary information is returned in `*stats` if this input is not NULL: `stats[0]` is assigned the geometric mean of $F_{H\alpha}$ (in W m$^{-2}$); `stats[1]` is assigned the standard deviation of ln $F_{H\alpha}$; `stats[2]` is assigned the skewness of ln $F_{H\alpha}$; and `stats[3]` is assigned the kurtosis of ln $F_{H\alpha}$.

### K. Routines to compute densities of observable galaxies: WL

```
double get_dNdA_WL_gal2(double reffmin, double reffmax, double zmin,
   double zmax, char GalaxyCat[], double lambda_min, double lambda_max,
   PSF_DATA *psf, int N_exp, double var_1exp, double calib_1exp, double
   max_ellip_err, double *dNeffdA, char OutFile[], char outmode);
double get_dNdA_WL_gal(double reffmin, double reffmax, double zmin,
   double zmax, char GalaxyCat[], double lambda_min, double lambda_max,
   PSF_DATA *psf, int N_exp, double var_1exp, double calib_1exp, double
   max_ellip_err, double *dNeffdA);
```

Either of these two functions will return the number density of galaxies (per deg$^2$) with measurable shapes in the range $z_{min} < z < z_{max}$ and $r_{eff,min} < r_{eff} < r_{eff,max}$. The galaxies are taken from the file name `GalaxyCat`. The routine requires ancillary data: the PSF, noise variance per exposure (in electrons$^2$ arcsec$^{-2}$), and the calibration (electrons per exposure for a 0 magnitude AB source), and number of exposures. If `dNeffdA` is not null, the effective number density of galaxies (per deg$^2$) is returned to `*dNeffdA`.

The difference between these two functions is that `get_dNdA_WL_gal2` outputs a table of resolved galaxies to `OutFile[]` (unless it receives the null pointer, in which case the behavior is identical to `get_dNdA_WL_gal` – indeed, the latter calls `get_dNdA_WL_gal2` with the null pointer). The output file is a 4-column list of galaxies: each row contains the galaxy ID; the redshift; the effective radius in arcsec; and the ellipticity uncertainty $\sigma_e$. The output file is either written ('w') or appended ('a') depending on `outmode`.

### L. Routines to compute the foreground radiation and absorption

```
double Galactic_Alambda__EBV(double lambda);
```

The Galactic reddening law, $A_\lambda/E(B-V)$, given λ in μm.

```
double get_zodi_bkgnd(double ecl_lat, double ecl_dlon, double
   lambda_min, double lambda_max, THROUGHPUT_DATA *T);
```

The zodiacal light intensity in photons/m$^2$/s/arcsec$^2$ between $\lambda_{min}$ and $\lambda_{max}$, observed through the throughput curve T. The ecliptic latitude β (`ecl_lat`) and longitude relative to Sun $\lambda - \lambda_\odot$ (`ecl_dlon`) are also required.

### M. Routines to compute noise and related quantities

```
double get_BB_flux_filter(THROUGHPUT_DATA *Thr, double lambda_min,
    double lambda_max, double T);
```

This routine computes the blackbody radiation flux $N$ (units: photons per $\mu m^2$ per s per sr) at temperature $T$ (in K) between wavelengths $\lambda_{min}$ and $\lambda_{max}$ (in μm) observed through a filter of throughput `Thr`.

```
double getThermalBkgndTel(THROUGHPUT_DATA *Thr, double lambda_min,
    double lambda_max, double filter_throughput, double Ttel, double
    fratio, int ismask, double centobs, double th_out_rad, double
    pmsm_emissivity);
```

This routine returns the thermal background rate in electrons per pixel per second from the telescope (primary and secondary mirrors, and associated black surfaces at the same temperature: spiders, obscurations and baffles, and radiation that overilluminates the primary mirror). It does <u>not</u> include margins, or emission from downstream components such as the tertiary mirror, fold flats, filters/prisms, cameras or reimaging optics, the cold mask, or detector enclosure – these are to be added separately.

The inputs to this routine are:

- `Thr`                 throughput table for telescope, excluding filter
- $\lambda_{min}$       minimum wavelength of filter (μm)
- $\lambda_{max}$       maximum wavelength of filter (μm)
- `filter_throughput`
                        filter throughput in band
- $T_{tel}$             telescope temperature (primary + secondary + associated structures) in Kelvin
- `fratio`              f/ratio of the telescope (open aperture only; increased by outer mask)
- `ismask`              flag: 0 = no cold mask, 1 = with cold mask
- `centobs`             linear central obscuration (only used w/o mask)
- `th_out_rad`          outer radius of thermal emission annulus divided by outer radius of primary mirror (>1; only used w/o mask)
- `pmsm_emissivity`
                        emissivity of PM + SM mirror surfaces

```
double getThermalBkgndTot(THROUGHPUT_DATA *Thr, double lambda_min,
    double lambda_max, double filter_throughput, double fratio, double
    centobs, THERMAL_DATA *thermal, int component);
```

This routine computes the total thermal background in the detector from all three sources (telescope; aft optics upstream of the filter; and the filter and downstream components). The inputs are:

- `Thr`                 throughput table for telescope, excluding filter
- $\lambda_{min}$       minimum wavelength of filter (μm)
- $\lambda_{max}$       maximum wavelength of filter (μm)

- `filter_throughput`
  filter throughput in band
- `fratio` f/ratio of the telescope (open aperture only; increased by outer mask)
- `centobs` linear central obscuration (only used w/o mask)
- `thermal` data structure with additional required information
- `component` an integer describing which source of thermal background is considered: 0 = all; 1 = telescope only; 2 = optics upstream of filter; 3 = filter and downstream components

This routine applies margin (given by the `THERMAL_MARGIN_FACTOR` macro) to the telescope and aft optics contributions; the downstream components contribution is assumed to carry its own margin.

```
double getNoiseTotal(double t_exp, double ct, double rnfloor, int
  mode);
```

This routine computes the noise variance in a pixel based on the exposure time `t_exp` (in seconds), the number of electrons counted during the exposure `ct` (including the dark current and any other sources with Poisson arrival times), the read noise floor `rnfloor` (in $e^-$ rms), and the fitting mode `mode`.
   The legal modes are currently:

| Mode | Description |
|---|---|
| 0 | Quadrature sum of read noise floor and Poisson noise. Appropriate for CCD data. WFIRST does not use CCDs but this option is provided for testing, and to enable future use of the ETC with e.g. Euclid. |
| 1 | Unweighted SUTR fit, with the read noise floor added in quadrature. |
| 2 | Unweighted SUTR fit, with the read noise floor replacing the read noise term if it falls below floor. |

Modes 1 and higher require the use of the `#defined` macros `T_FRAME` (the time in seconds to read out an entire frame; this is 1.3 for a 2k×2k detector and 32 channel readout at 100 kHz each) and `VAR_READ_FRAME` (the variance of the read noise per frame; e.g. 200 for statistical read noise of 20 $e^-$ rms per CDS).

### N. <u>I/O and associated routines</u>

```
void error_check_thermal(THERMAL_DATA *thermal);
```

This routine checks a `THERMAL_DATA` structure for unusual or illegal values. The program exits if an illegal value is found, and issues a warning if unusual values are found.

```
void read_thermal_data(char Data[], THERMAL_DATA *thermal);
```

This routine obtains thermal data from a string (this is generally read from a configuration file, but this routine handles only the string). It reads a string `Data[]` and stores the associated information in the structure `thermal`. It then checks for errors and warnings using `error_check_thermal`.

# 9. Version History

Version 1:  March 19, 2011
First version, with basic BAO functionality only (including a single HαLF model and constant throughput).

Version 2:  March 21, 2011
Added support for telescope configuration files, including separate all-order and 1st order throughputs.

Version 3:  March 27, 2011
Improved speed by eliminating redundant calls to MTF calculator.
Added additional outputs.
Fixed bug in output for kurtosis of ln $F_{H\alpha}$.

Version 4:  April 3, 2011
Added new HαLF models.
Fixed bug in implementation of Geach *et al* HαLF (minor: up to 4% change in $L_*$).

Version 5:  April 9, 2011
Added WL calculator including source galaxy catalog input/processing, penalty factor function, and polychromatic PSFs.
Improved accuracy in **u** integration for $\Omega_{eff}$ calculation.

Version 6:  April 11, 2011
Added support for aberrations.
Added filter throughput.
Restricted range in **u** integration for BAO $\Omega_{eff}$ calculation.
Updated cosmology to agree with FoMSWG/WMAP5 to more decimal places.

Version 7:  May 4, 2011
Included code to catch incompatible compilation options.
Introduced –DOUT_WL_CAT and –DWFE_OVERRIDE options.
Added independent SNR cut for WL source galaxies.
Added finite line width smearing in BAO mode.
Added more realistic model for noise variance on NIR detectors (v6 and previous had behavior equivalent to `mode==0` in `getNoiseTotal`, which was too optimistic).
Added thermal background model.

Version 8:　　　　October 1, 2011
Added PZCAL mode.
Added updated CMC galaxy size distribution model ($j$=2), and correspondingly changed the suggested command line input.
Changed `CCD_MODE` behavior to correspond to pixel parameters appropriate for a CCD rather than HgCdTe detector.
Changed `WFE_OVERRIDE` behavior to switch off WL-specific inputs that are not used in this mode.
Fixed bug in $n_{\rm eff}$ computation (v7 and previous were too pessimistic; typical changes are ~3%).

Version 9:　　　　March 7, 2012
Changed detector options from macros to global variables set at the beginning of the run; allows user to choose detector without recompiling.
Added `-DWFE_HOPKINS_EXP` flag for alternative wavefront distribution.
Added support for completeness in the BAO mode.
Fixed bugs in error reporting.

Version 10:　　　　March 28, 2012
Added the Sobral *et al* HαLF option.
Added output on photon counts per unit line flux in the BAO mode.
Added source continuum to the noise model.
Switched default BAO cuts to SNRo instead of SNRe (the latter is available as the `-DUSE_SNRE option`).

## 10.   Acknowledgements

The development of this ETC has benefited from comparisons with other codes and/or calculations by Ed Cheng and Michael Levi.
Peter Capak, Ed Cheng, Dave Content, Cliff Jackson, and David Sobral have been very helpful in providing input models, comments, and feedback.
During the development of the ETC, Christopher Hirata has been supported by the U.S. Department of Energy (DOE-FG03-92-ER40701 and DOE.DE-SC0006624), the National Science Foundation (NSF AST-0807337), and the David & Lucile Packard Foundation.

## 11.   Acronyms

BAO　　　　Baryon Acoustic Oscillations
CCD　　　　Charge Coupled Device
CDS　　　　Correlated Double Sample
CMC　　　　COSMOS Mock Catalog
EE50　　　　50% Encircled Energy radius
ETC　　　　Exposure Time Calculator

| | |
|---|---|
| EW | Equivalent Width |
| FoMSWG | Figure of Merit Science Working Group (2008) |
| FWHM | Full Width at Half Maximum |
| HαLF | Hydrogen-α Luminosity Function |
| IDRM | Interim Design Reference Mission (2011) |
| ISWG | Interim Science Working Group (2009—10) |
| JDEM | Joint Dark Energy Mission |
| JDEMΩ | Joint Dark Energy Mission, Ω configuration |
| MDLF | Minimum Detectable Line Flux |
| MTF | Modulation Transfer Function |
| NIR | Near InfraRed |
| PSF | Point Spread Function |
| PZCAL | Photometric Z (redshift) CALibration |
| RMSWFE | Root Mean Square Wave Front Error |
| SDSS | Sloan Digital Sky Survey |
| SED | Spectral Energy Distribution |
| SNAP | Supernova Acceleration Probe (a JDEM concept) |
| SNR | Signal-to-Noise Ratio |
| SNRe | Signal-to-Noise Ratio – Expected |
| SNRo | Signal-to-Noise Ratio – Observed |
| SUTR | Sample Up The Ramp |
| TRL | Technology Readiness Level |
| WFIRST | Wide Field InfraRed Space Telescope |
| WL | Weak Lensing |
| WMAP | Wilkinson Microwave Anisotropy Probe |